\documentclass[12pt,twoside,a4paper]{article}
\usepackage{amsmath,amssymb,latexsym,theorem,natbib,epsfig,color,subfigure}
\usepackage{multirow,graphicx,array,rotating}  
\usepackage{epstopdf,afterpage,wasysym}
\usepackage{verbatim}
\usepackage[normalem]{ulem}
\newcommand{\proofend}{\hfill$\square$}

\newfont{\msa}{msam10 scaled\magstep1}
\newfont{\ssmsa}{msam9}

\def\crps{\mathop{\hbox{\rm CRPS}}}
\def\crpss{\mathop{\hbox{\rm CRPSS}}}

\def\twCRPS{\mathop{\hbox{\rm twCRPS}}}
\def\twCRPSS{\mathop{\hbox{\rm twCRPSS}}}
\def\md{\mathrm{MD}}
\def\ei{\mathrm {Ei}}

\numberwithin{equation}{section}

\title{Truncated generalized extreme value distribution based EMOS model for calibration of wind speed ensemble forecasts}

\author{{\sc S\'andor Baran$^{1}$},  {\sc Patr\'\i cia Szokol$^{1}$} and {\sc Marianna Szab\'o$^{1,2}$} \vspace*{0.5cm}\\
         $^1$Faculty of Informatics, University of Debrecen\\
         $^2$Doctoral School of Informatics, University of Debrecen\\
         Kassai \'ut 26, H-4028 Debrecen, Hungary
         }

        \date{}

\begin{document}
\pagestyle{myheadings}

\maketitle

\begin{abstract}
In recent years, ensemble weather forecasting have become a routine at all major weather prediction centres. These forecasts are obtained from multiple runs of numerical weather prediction models with different initial conditions or model parametrizations. However, ensemble forecasts can often be underdispersive and also biased, so some kind of post-processing is needed to account for these deficiencies. One of the most popular state of the art statistical post-processing techniques is the ensemble model output statistics (EMOS), which provides a full predictive distribution of the studied weather quantity.

We propose a novel EMOS model for calibrating wind speed ensemble forecasts, where the predictive distribution is a generalized extreme value (GEV) distribution left truncated at zero (TGEV). The truncation corrects the disadvantage of the GEV distribution based EMOS models of occasionally predicting negative wind speed values, without affecting its favorable properties. The new model is tested on four data sets of wind speed ensemble forecasts provided by three different ensemble prediction systems, covering various geographical domains and time periods. The forecast skill of the TGEV EMOS model is compared with the predictive performance of the truncated normal, log-normal and GEV methods and the raw and climatological forecasts as well. The results verify the advantageous properties of the novel TGEV EMOS approach. 

\end{abstract}

\section{Introduction}
\label{sec:sec1}
Wind speed has become one of the most important weather quantities in our rapidly changing economy, hence precise and reliable wind forecasting is of utmost importance in renewable energy production or in air pollution modelling. At the base of forecasting such - and many other - weather variables lie the calculations of numerical weather prediction (NWP) models, which rely on the physical and chemical models of the atmosphere and the oceans.
Accounting for the uncertainties of the process and the sometimes unreliable initial conditions it is customary to run multiple instances of the NWP models with its initial conditions perturbed. The resulting system is called an ensemble of forecasts \citep{lei74}, and it provides the possibility of probabilistic forecasting \citep{gr05}, where together with the forecasts the corresponding information about forecast uncertainty is also estimated. However, as has been observed with several operational ensemble prediction systems (EPSs), ensemble forecasts often suffer from systematic errors such as bias or lack of calibration, which problems need to be accounted for \citep[see e.g.][]{bhtp05}.  A popular approach is to use some form of statistical post-processing \citep{buizza18}.

In the last decades various statistical calibration methods have been developed for a wide range of weather quantities including parametric models providing full predictive distributions \citep{rgbp05,grwg05}, non-parametric approaches \citep[see e.g.][]{fh07,brem19} or most recently, machine learning techniques \citep{rl18,tm20}. This paper focuses on parametric post-processing where one of the most widely used methods is the ensemble model output statistics (EMOS) suggested by \citet{grwg05}. It fits a single probability distribution to the ensemble forecast with its parameters depending on the ensemble members. Different weather quantities require different probability laws as predictive distributions, moreover, the link functions connecting the parameters of these distributions to the ensemble members might also differ. E.g. a normal distribution provides a reasonable model for temperature and pressure \citep{grwg05}, whereas for the non-negative and skew distributed wind speed, according to \citet{tg10}, a truncated normal (TN) distribution makes a good choice. In order to provide a better fit to high wind speed values, \citet{lt13} and \citet{bl15} suggest models based on generalized extreme value (GEV) and log-normal (LN) distributions, respectively, and a regime-switching approach combining the advantages of these heavy tailed laws with those of the light tailed TN model. More flexibility can be obtained by mixture EMOS models combining light and heavy tailed distributions, where the parameters and weights of a mixture of two forecast laws are estimated jointly \citep{bl16}. However, a general disadvantage of these latter approaches is the increased computation cost. A more general approach to improving forecast skill is based on a two-step combination of predictive distributions from individual post-processing models. In the first step, individual EMOS models based on single parametric distributions are estimated, whereas in the second step the forecast distributions are combined utilizing state of the art forecast combination techniques \citep[see e.g.][]{gr13,bcr18,bl18}.

In the present work we concentrate on EMOS models based on a single parametric distribution. The case studies of \cite{lt13} and \cite{bl15} revealed the superiority of the GEV EMOS model compared with the competing TN and LN EMOS approaches, especially for high wind speeds. However, the GEV model has the disadvantage of assigning positive probability to negative wind speed values. We propose a novel EMOS approach to calibrating wind speed ensemble forecasts, where the predictive distribution is a left truncated GEV distribution with cut-off at \ $0$ \ (TGEV).  On the basis of four case studies using wind speed forecasts of three different EPSs, the forecast skill of the TGEV EMOS model is compared with the predictive performance of the TN, LN and GEV EMOS models, the climatological forecasts and the raw ensemble as well.

The paper is organized as follows. Section \ref{sec:Data} contains the detailed description of the four wind speed data sets. In Section \ref{sec:EMOS} the applied EMOS models, including the novel TGEV EMOS approach, are reviewed, and the methods of parameter estimation and model verification are given. The results of the four case studies are provided in  Section \ref{sec:Res}, followed by a concluding Section \ref{sec:Conc}. Finally, details of calculations and some additional results are given in the Appendix.

\section{Data}
\label{sec:Data}
In order to provide a fair comparison with the existing distribution-based EMOS models, first we consider the same three data sets of ensemble forecasts and corresponding observations as in \citet{bl15} (and later studied in \citet{bl16,bl18}), which differ in the observed wind quantity, in the forecast lead time and in the stochastic properties of the ensemble. Each data set contains ensemble predictions for a single forecast horizon ranging from 24 h to 48 h, hence we call them short-range forecasts.
For these data we limit the description to a short summary and refer to \citet{bl15} and the references therein for more details. 
Further, we compare the predictive performance of the different EMOS models on a much larger data base, providing ensemble forecasts with different lead times ranging up to 360 h.

\subsection{Short-range ensemble forecasts}
\label{subs:subs2.1}

\subsubsection{UWME forecasts}
\label{subs:subs2.1.1}
The eight members of the University of Washington mesoscale ensemble (UWME) are generated by separate runs of the fifth generation Pennsylvania State University-National Center for Atmospheric Research mesoscale model (PSU-NCAR MM5) with different initial conditions \citep{grell95}. The EPS domain covers the Pacific Northwest region of North America with a 12 km grid and the data set at hand contains 48 h ahead forecasts and the corresponding validating observations of the 10 m maximal wind speed (maximum
of the hourly instantaneous wind speeds over the previous 12 h, given in m/s, see e.g. \citet{sgr10}) for 152 stations in the Automated Surface Observing Network \citep{nws98} in the U.S. states of Washington, Oregon, Idaho, California and Nevada for calendar years 2007--2008. The forecasts are initialized at 0000 UTC and the generation of the ensemble ensures that its members are clearly distinguishable. Our analysis is focused on calendar year 2008 with additional data from December 2007 used for model training.  Removing days and locations with missing data and stations where data are only available on a very few days results in 101 stations with a total of 27\,481 individual forecast cases.

\subsubsection{ALADIN-HUNEPS ensemble}
  \label{subs:subs2.1.2}
  The Aire Limit\'ee Adaptation dynamique D\'eveloppement International-Hungary Ensemble Prediction System (ALADIN-HUNEPS) of the Hungarian Meteorological Service (HMS) covers a large part of continental Europe with a horizontal resolution of 8 km. The forecasts are obtained by dynamical downscaling of the global ARPEGE\footnote{Action de Recherche Petite Echelle Grande Echelle}-based PEARP\footnote{Pr\'evisino d'Ensemble ARPEGE} system of Met\'eo-France \citep{hkkr06,dljbac}. The EPS provides one control member obtained from the unperturbed analysis and 10 members calculated using perturbed initial conditions. These members are statistically indistinguishable and thus can be considered as exchangeable, which fact should be taken into account in the formulation of post-processing models. We use ensembles of 42 h ahead forecasts (initialized at 1800 UTC) of the 10 m instantaneous wind speed (in m/s) issued for 10 major cities in Hungary for the one-year period  1 April 2012 -- 31 March 2013, together with the corresponding validation observations. 6 days with missing forecasts and/or observations are excluded from the analysis.

\subsubsection{ECMWF ensemble forecasts for Germany}
  \label{subs:subs2.1.3}
The operational EPS of the European Centre for Medium-Range Weather Forecasts (ECMWF) comprises 50 perturbed (thus exchangeable) members and operates on a global 18 km grid \citep{mbp96,lp08}. First we consider 24 h ahead ECMWF forecasts of 10 m daily maximum wind speed initialized at 0000 UTC for the period between 1 February 2010 and 30 April 2011 along with corresponding verifying observations of 228 synoptic observation (SYNOP) stations over Germany. This data set is identical to the one studied in \citet{lt13} and in \citet{bl15,bl16}. Post-processed forecasts are verified on the one-year period between 1 May 2010 and 30 April 2011 containing 83\,220 individual forecast cases, whereas forecast-observation pairs from April 2010 are used for training purposes.

\subsection{Global ECMWF forecasts with different forecast horizons}
\label{subs:subs2.2}
In order to compare the predictive performance of the various EMOS models for different prediction horizons, we also investigate a global data set of ECMWF ensemble forecasts of 10 m daily maximal wind speed with lead times from 1 day up until 15 days initialized at 1200 UTC between 1 January 2014 and 24 June 2018, and validating SYNOP observations for calendar years 2014--2018. Thus, one has observations and corresponding ensemble forecasts with 15 different lead times for the period 16 January 2014 -- 25 June 2018 with the exception of two days in between with missing forecast data. For the sake of consistency our analysis is restricted to SYNOP stations with complete data, meaning 1059 stations in Europe and Asia.

\section{Ensemble model output statistics}
\label{sec:EMOS}
As already mentioned in the Introduction, EMOS is a commonly used method of statistical post-processing, which fits a single probability distribution to the ensemble forecast with parameters depending on the ensemble members. In what follows, let \ $f_1, f_2, \ldots , f_K$ \ denote a wind speed ensemble forecast for a given location, time and lead time under the assumption that the ensemble members are not exchangeable, so the individual members can be clearly distinguished and tracked either based on the individual initial conditions, or as one can depict a systematic behaviour of each ensemble member. Examples of EPSs with non-exchangeable members are the UWME introduced in Section \ref{subs:subs2.1.1} or the 30-member Consortium for Small-scale Modelling EPS of the German Meteorological Service \citep{btg13}.  

However, recently most operational EPSs incorporate ensembles where at least some members are generated using perturbed initial conditions. Such groups of exchangeable forecasts appear e.g. in the ALADIN-HUNEPS ensemble and in the operational ECMWF ensemble described in Sections \ref{subs:subs2.1.2} and \ref{subs:subs2.1.3}, respectively, but one can also mention multi-model EPSs such as the Grand Limited Area Model Ensemble Prediction System ensemble \citep{iversen11}. In the following sections, if we have \ $M$ \ ensemble members divided into \ $K$ \ exchangeable groups, where the $k$th group contains \ $M_k\geq 1$ \ ensemble members \ $(\sum_{k=1}^K M_k=M)$, \ then notation \ $\overline f_k$ \ will be used for the mean of the corresponding $k$th ensemble group. Further, the overall ensemble mean and variance will be denoted by \ $\overline f$ \ and \ $S^2$, \ respectively. 

\subsection{EMOS models for wind speed}
\label{subs:subs3.1}
To model wind speed a non-negative and skewed distribution is required, such as Weibull \citep{jhmg78} or gamma \citep{gtpd98} laws. Gamma distribution also serves as underlying law in a Bayesian model averaging \citep{sgr10} approach to parametric post-processing of wind speed ensemble forecasts, whereas in EMOS modelling truncated normal (TN), log-normal (LN) and generalized extreme value (GEV) distributions have been utilized so far. Note, that TN and LN EMOS models have already been implemented in the {\tt ensembleMOS} package of {\tt R} \citep{emos}.

\subsubsection{Truncated normal EMOS model}
\label{subs:subs3.1.1}
Starting with the fundamental work of \citet{tg10}, TN distribution became a popular base for EMOS predictive distributions of wind speed \citep[see e.g.][]{lb17,brem19}. Denote by \ $\mathcal{N}_0(\mu,\sigma^2)$ \ the TN distribution  with location \ $\mu$, \ scale \ $\sigma > 0$, \ and lower truncation at $0$, having probability density function (PDF)
\begin{equation*}
    g(x|\mu,\sigma) := 
    \frac1{\sigma}\varphi\big((x-\mu)/\sigma\big) / \Phi(\mu/\sigma), \qquad \text{if \ $x\geq 0$,}
\end{equation*}
and \  $g(x|\mu,\sigma):=0$, \ otherwise, where \ $\varphi$ \ is the PDF, while \ $\Phi$ \ denotes the cumulative distribution function (CDF) of the standard normal distribution. For the TN EMOS predictive distribution the location and  scale are linked to the ensemble members via equations 
 \begin{equation}
 \label{eq:tnlocscale}
\mu = a_0 + a_1f_1 + \cdots + a_Kf_K \qquad \text{and}\qquad \sigma^2 = b_0 + b_1 S^2.
\end{equation}
where \ $a_0\in {\mathbb R}$ \ and \ $a_1,\ldots ,a_K,b_0,b_1\geq 0$. \

If the ensemble can be split into \ $K$ \ groups of exchangeable members, then forecasts within a given group will share the same location parameter \citep{gneiting14, wilks18} resulting in link functions 
 \begin{equation}
 \label{eq:tnlocscaleEx}
\mu = a_0 + a_1\overline f_1 + \cdots + a_K\overline f_K \qquad \text{and}\qquad \sigma^2 = b_0 + b_1 S^2.
\end{equation}
According to the optimum score estimation principle of \citet{gr07}, model parameters \ $a_0, a_1, \ldots ,a_K$ \ and \ $b_0,b_1$ \ are estimated by optimizing the mean value of a proper verification score over the training data, see Section \ref{subs:subs3.2}.

\subsubsection{Log-normal EMOS model}
\label{subs:subs3.1.2}
To address the modelling of large wind speeds \citet{bl15} propose an EMOS approach based on an LN distribution. This distribution is more applicable for high wind speed values due to its heavier upper tail.  The PDF of the LN distribution \ $\mathcal{LN}(\mu,\sigma)$ \ with parameters \ $\mu$ \ and  \ $\sigma > 0$ \ is
\begin{equation*}
    h(x|\mu,\sigma) := 
    \frac{1}{x\sigma}\varphi\big((\log x -\mu)/\sigma\big), \qquad \text{if \ $x\geq 0$,}
\end{equation*}
and \ $ h(x|\mu,\sigma) := 0$, \ otherwise, while the mean \ $m$ \ and variance \ $v$ \ are
\begin{equation*}
m = {\mathrm e}^{\mu+\sigma^2 /2} \qquad \text{and} \qquad v = {\mathrm e}^{2\mu+\sigma^2}\big({\mathrm e}^{\sigma^2}-1\big),
\end{equation*}
respectively. Obviously, an LN distribution can also be parametrized by these latter two quantities via equations
\begin{equation*}
\mu = \log\left(\frac{m^2}{\sqrt{v+m^2}}\right) \qquad \text{and} \qquad \sigma = \sqrt{\log\left( 1 + \frac{v}{m^2} \right)},
\end{equation*}
and in the LN EMOS model of \citet{bl15}  \ $m$ \ and \ $v$ \ are affine functions of the ensemble and the ensemble variance, respectively, that is
 \begin{equation}
 \label{eq:lnlocscale}
 m = \alpha_0 + \alpha_1f_1 + \cdots + \alpha_Kf_K \qquad \text{and}\qquad  v = \beta_0 + \beta_1S^2.
 \end{equation}
To estimate mean parameters \ $\alpha_0 \in{\mathbb R}, \alpha_1, \ldots, \alpha_K\geq 0$ \ and variance parameters \ $\beta_0,\beta_1 \geq 0$, one can again use the optimum score estimation principle and minimize an appropriate verification score over the training data.

In the case of existence of groups of exchangeable ensemble members, similar to \eqref{eq:tnlocscaleEx}, the equation for the mean in \eqref{eq:lnlocscale} is replaced by
 \begin{equation}
 \label{eq:lnlocscaleEx}
 m = \alpha_0 + \alpha_1\overline f_1 + \cdots + \alpha_K\overline f_K.
 \end{equation}
 
\subsubsection{Generalized extreme value distribution based EMOS models}
\label{subs:subs3.1.3}
As an alternative to the TN EMOS approach exhibiting good predictive performance for high wind speed values, one can consider the EMOS model of \citet{lt13} based on a generalized extreme value distribution \ $\mathcal{GEV}(\mu,\sigma,\xi)$ \ with location \ $\mu$, \ scale \ $\sigma >0$ \ and shape \ $\xi$ \ defined by CDF
\begin{equation}
\label{eq:gevCDF}
G(x|\mu,\sigma,\xi) := \begin{cases}
\exp\Big(-\big[1+\xi(\frac{x-\mu}{\sigma})\big]^{-1/\xi}\Big), & \quad \text{if \ $\xi \neq 0$;}\\
\exp\Big(-\exp\big(-\frac{x-\mu}{\sigma}\big)\Big),  & \quad \text{if \ $\xi = 0$},
\end{cases}
\end{equation}
for \ $1+\xi(\frac{x-\mu}{\sigma}) > 0$ \ and \ $G(x|\mu,\sigma,\xi) := 0$, \ otherwise.

The model proposed by \citet{lt13} uses  location and scale parameters
\begin{equation}
\label{eq:gevlocscale}
\mu = \gamma_0 + \gamma_1f_1 + \cdots + \gamma_Kf_K \qquad \text{and} \qquad \sigma = \sigma_0 + \sigma_1\overline{f},
\end{equation}
with \ $\sigma_0,\sigma_1\geq 0$, \  while the shape parameter \ $\xi$ \ does not depend on the ensemble members.

However, as argued in \citet{lt13} and in \citet{bl15}, the GEV EMOS model has the disadvantage of forecasting negative wind speed with a positive probability. As a solution we propose a novel EMOS model where the predictive GEV distribution is truncated from below at \ $0$. \ For \ $x\geq 0$ \ the CDF of this truncated GEV (TGEV) distribution \ $\mathcal{TGEV}(\mu,\sigma,\xi)$ \ with location \ $\mu$, \ scale \ $\sigma >0$ \ and shape \ $\xi$ \  equals
\begin{equation}
    \label{eq:tgevCDF}
    G_0(x|\mu,\sigma,\xi)=\begin{cases}
\frac{G(x|\mu,\sigma,\xi)-G(0|\mu,\sigma,\xi)}{1-G(0|\mu,\sigma,\xi)}, & \text{if \ $G(0|\mu,\sigma,\xi) < 1$};\\
1,  & \text{if \ $G(0|\mu,\sigma,\xi) = 1$},
\end{cases} 
\end{equation}
whereas negative values are obviously excluded from the support set of the TGEV distribution. For \ $\xi <1$ \ (and \  $G(0|\mu,\sigma,\xi) < 1$) \ the \ $\mathcal{TGEV}(\mu,\sigma,\xi)$ \ distribution has a finite mean of
\begin{equation}
   \label{eq:tgevMean}
    \begin{cases}
    \mu + \big(\Gamma(1-\xi)-1\big)\frac{\sigma}{\xi}, & \quad \text{if \ $\xi> 0$ \ and \ $\xi\mu - \sigma >0$;}\\
    \mu - \frac{\sigma}{\xi}+ \frac{\sigma(\Gamma_{\ell}(1-\xi,[1-{\xi\mu}/{\sigma}]^{-1/\xi}))/\xi }{1-\exp (-[1-{\xi\mu}/{\sigma}]^{-1/\xi})}, & \quad \text{if \ $\xi \ne 0$ \ and \ $\xi\mu - \sigma \leq 0$;}\\
    \frac{\mu + \sigma(C-\ei(-\exp[{\mu}/{\sigma}]))}{1-\exp(-\exp[\mu/\sigma])}, & \quad \text{if \ $\xi= 0$,}
    \end{cases}
\end{equation}
where \ $\Gamma$ \ and \ $\Gamma_{\ell}$ \ denote the gamma and the lower incomplete gamma function, respectively, defined as
\begin{equation*}
    \Gamma(a)=\int_0^{\infty} t^{a-1}{\mathrm e}^{-t} {\mathrm d} t \qquad \text{and} \qquad
    \Gamma_{\ell}(a,x)=\int_0^x t^{a-1}{\mathrm e}^{-t} {\mathrm d} t,
\end{equation*} 
and \ $\ei(x)$ \ is the exponential integral
\begin{equation*}
\ei(x)=\int_{-\infty}^{x}\frac{e^t}{t}\mathrm{d}t=C+\ln|x|+\sum\limits_{k=1}^{\infty} \frac{x^k}{k!k}
\end{equation*}
with \ $C$ \ being the Euler–Mascheroni constant. It is important to emphasize, that the case \ $\xi<0$ \ and \ $\xi\mu-\sigma>0$, \ does not appear in the formula \eqref{eq:tgevMean}, since in that case the PDF of \ $\mathcal{GEV}(\mu,\sigma,\xi)$ \ is positive only on \ $]-\infty,\mu-\sigma/\xi]\subset\mathbb{R}_-$. \ Further, as for \ $\xi>0$ \ and \ $\xi\mu-\sigma>0$ \ the support of \ $\mathcal{GEV}(\mu,\sigma,\xi)$ \ is \ $[\mu-\sigma/\xi,\infty[\subset\mathbb{R}_+$, \ truncation does not change the distribution and the means of \  $\mathcal{GEV}(\mu,\sigma,\xi)$ \ and \ $\mathcal{TGEV}(\mu,\sigma,\xi)$ \ distributions coincide. For the proof of the remaining two cases of \eqref{eq:tgevMean} see Appendix \ref{appendix_mean}.

The parameters of the TGEV EMOS model are also linked to the ensemble members according to \eqref{eq:gevlocscale}, which is replaced by 
\begin{equation}
\label{eq:gevlocscaleEx}
\mu = \gamma_0 + \gamma_1\overline f_1 + \cdots + \gamma_K\overline f_K \qquad \text{and} \qquad \sigma = \sigma_0 + \sigma_1\overline{f},
\end{equation}
in the exchangeable case. Note that alternative expressions
\begin{equation*}
  \sigma = \sigma_0 + \sigma_1 S, \qquad \sigma = \sqrt{\sigma_0 + \sigma_1 S^2} \qquad \text{and} \qquad \sigma = \sigma_0 + \sigma_1 \md
\end{equation*}
of the scale have also been tested, where
\begin{equation*}
  \md := \frac 1{K^2}\sum_{k,\ell =1}^K \big|f_k - f_{\ell}\big|
\end{equation*}
is the ensemble mean absolute difference \citep[see e.g.][]{sch14,bbpbb20}. However, in our case studies TGEV EMOS models with link functions \eqref{eq:gevlocscale} and \eqref{eq:gevlocscaleEx} show the best predictive performance.

\subsection{Training data selection and verification scores}
  \label{subs:subs3.2}
As mentioned before, estimates of the unknown parameters of the EMOS models described in Sections \ref{subs:subs3.1.1} -- \ref{subs:subs3.1.3} can be obtained with the help of the optimum score estimation principle of \citet{gr07}, that is by optimizing a proper scoring rule over an appropriately chosen training data set. Here we consider the standard approach in EMOS modelling and use rolling training periods. This means that model parameters for a given date are obtained using ensemble forecasts and corresponding validating observations for the preceding \ $n$ \ calendar days. Given a training period length, there are two traditional approaches to spatial selection of training data \citep{tg10}. The global (or regional) approach uses ensemble forecasts and validating observations from all available stations during the rolling training period resulting in a single set of parameters for the whole ensemble domain. By contrast, the local estimation produces distinct parameter estimates for different stations by using only the training data of the given station. Local models typically result in better predictive performance compared with regional models \citep[see e.g.][]{tg10}; however, require significantly longer training periods to avoid numerical stability issues \citep{lb17}. In the case studies of Section \ref{sec:Res} examples of both estimation techniques are shown.

In atmospheric sciences the most popular scoring rules are the logarithmic score \citep[LogS;][]{good52} and the continuous ranked probability score \citep[CRPS; see e.g.][]{w11}. The former is the negative logarithm of the predictive PDF evaluated at the verifying observation, whereas for a (predictive) CDF \ $F$ \ and real value (verifying observation) \ $x$ \ the latter is defined as 
\begin{equation}
    \label{eq:CRPSdef}
\crps(F,x) := \int_{-\infty}^{\infty}\Big[F(y)-{\mathbb I}_{\{y\geq x\}}\Big]^2{\mathrm d}y ={\mathsf E}|X-x|-\frac 12
{\mathsf E}|X-X'|,
\end{equation}
where \ $X$ \ and \ $X'$ \ are independent random variables distributed according to \ $F$ \ and having a finite first moment, while \ ${\mathbb I}_H$ \ denotes the indicator function of a set \ $H$. \ Note that both LogS and CRPS are negative oriented scores, that is the smaller the better. Further, the optimization with respect to the logarithmic score results in the maximum likelihood (ML) estimation of the parameters, while the second expression in \eqref{eq:CRPSdef} implies that the CRPS can be expressed in the same unit as the observation. For all wind speed models of Sections \ref{subs:subs3.1.1} -- \ref{subs:subs3.1.3} the CRPS can be expressed in closed form allowing efficient optimization procedures; for TN, LN and GEV laws we refer to \citet{tg10}, \citet{bl15} and \citet{ft12}, respectively. The CRPS of a TGEV distribution \ $\mathcal{TGEV}(\mu,\sigma,\xi)$ \ with CDF \ $G_0(x)$ \ derived from a GEV CDF \ $G(x)$ \ equals 
\begin{align}
  \label{eq:CRPStgev1}
\crps(G_0,x)=\big(2G_0(x)&\!-\!1\big)\Big(x\!-\!\mu\!+\!\frac{\sigma}{\xi}\Big)+
    \frac{\sigma}{\xi(1\!-\!G(0))^2}\Big[-2^{\xi}\Gamma_{\ell}\big(1-\xi,-2\ln G(0)\big) \\
    & +2G(0)\Gamma_{\ell}\big(1-\xi,-\ln G(0)\big)+2\big(1\!-\!G(0)\big)\Gamma_{\ell}\big(1-\xi,-\ln G(x)\big) \Big] \nonumber
\end{align}
for \ $\xi\ne 0$, \ whereas for \ $\xi=0$ \ we have
\begin{align}
  \label{eq:CRPStgev2}
    \crps(G_0,x)=&\,(x-\mu)\big(2G_0(x)-1\big) +\frac{\sigma}{(1\!-\!G(0))^2} \\ \nonumber
    &\times\Big(C-\ln 2+
    \ei\big(2\ln G(0)\big)+(G(0))^2\ln\big[-\ln G(0)\big]-
    2G(0)\ei\big(\ln G(0)\big)\Big)\\ \nonumber
    &+\frac{2\sigma}{1\!-\!G(0)}\Big[G(x)\ln\big[-\ln G(x)\big]-\ei\big(\ln G(x)\big) \Big].
\end{align}
For the proof of \eqref{eq:CRPStgev1} and \eqref{eq:CRPStgev2} see Appendix \ref{appendix_crps}.

In order to compare the predictive performance of the EMOS models for high wind speed values we also consider the threshold-weighted continuous ranked probability score \citep[twCRPS;][]{gr11}
\begin{equation}
    \label{eq:twCRPSdef}
\twCRPS(F,x) := \int_{-\infty}^{\infty}\Big[F(y)-\mathbb I_{\{y\geq x\}}\Big]^2 \omega(y) {\mathrm d}y,
\end{equation}
where \ $\omega(y)\geq 0$ \ is a weight function. Setting \ $\omega(y)\equiv 1$ \ results in the traditional CRPS (\ref{eq:CRPSdef}), whereas with the help of \ $\omega(y)=\mathbb I_{\{y\geq r\}}$ \ one can address wind speeds above a given threshold \ $r$. \ Note that in the case studies of Section \ref{sec:Res} the thresholds correspond approximately to the $90$th, $95$th and $98$th percentiles of the wind speed observations of all considered stations.

The improvement in terms of CRPS and twCRPS for a forecast \ $F$ \ with respect to a reference forecast \ $F_{ref}$ \ can be quantified using the continuous ranked probability skill score \citep[CRPSS; see e.g.][]{murphy73,gr07} and the threshold-weighted continuous ranked probability skill score \citep[twCRPSS;][]{lt13} 
\begin{equation*}
\crpss := 1-\frac{\overline{\crps}}{\overline{\crps}_{ref}} \qquad \text{and} \qquad \twCRPSS := 1-\frac{\overline{\twCRPS}}{\overline{\twCRPS}_{ref}},
\end{equation*}
respectively, where \ $\overline{\crps}$, \ $\overline{\twCRPS}$ \ and \ $\overline{\crps}_{ref}$, \ $\overline{\twCRPS}_{ref}$ \ denote the mean score values corresponding to \ $F$ \ and \ $F_{ref}$ \ over the verification data. Skill scores are obviously positively oriented, that is larger skill scores mean better predictive performance.

Point forecasts such as EMOS and ensemble medians and means can be evaluated using the mean absolute errors (MAEs) and the root mean squared errors (RMSEs), where the former is optimal for the median, whereas the latter is optimal for the mean forecasts \citep{gneiting11}.

The uncertainty in the verification scores is assessed with the help of confidence intervals for mean score values and skill scores. These intervals are calculated from 2 000 block bootstrap samples, which are obtained using the stationary bootstrap scheme with mean block length computed according to \citet{pr94}.

Simple and widely used tools of graphically assessing the calibration of probabilistic forecasts are the verification rank histogram (or Talagrand diagram) of ensemble predictions and its continuous counterpart, the probability integral transform (PIT) histogram. The verification rank is the rank of the verifying observation with respect to the corresponding ensemble forecast \citep[see e.g.][Section 8.7.2]{w11}, whereas the PIT is the value of the predictive CDF evaluated at the verifying observation  \citep{dawid84,rgbp05}. In the case of a properly calibrated $K$-member ensemble the verification ranks follow a uniform distribution on \ $\{1, 2, \ldots , K + 1\}$, \ while PIT values of calibrated predictive distributions are uniformly distributed on the \ $[0,1]$ \ interval.

Finally, calibration and sharpness of a predictive distribution can also be investigated by examining the coverage and average width of the \ $(1-\alpha)100\,\%, \ \alpha \in ]0,1[,$ \ central prediction interval, respectively. Here the coverage is the proportion of the validating observations located between the lower and upper \ $\alpha/2$ \ quantiles of the predictive CDF, and level \ $\alpha$ \ should be chosen to match the nominal coverage of the raw ensemble, that is \ $(K - 1)/(K + 1)100\,\%$, \ where again,  \ $K$ \ is the ensemble size. As the coverage of a calibrated predictive distribution should be around \ $(1-\alpha)100\,\%$, \  such a choice of \ $\alpha$ \ allows a direct comparison with the ensemble coverage.

\section{Results}
  \label{sec:Res}
  The forecast skill of the novel TGEV EMOS model proposed in Section \ref{subs:subs3.1.3}  is tested both on short-range (24 -- 48 h) wind speed forecasts of the 8-member UWME, of the 11-member ALADIN-HUNEPS ensemble and of the 50-member ECMWF ensemble, and on more recent global surface wind forecasts of the operational EPS of the ECMWF with lead times 1,2, \ldots ,15 days, for more details see Section \ref{sec:Data}. As reference models we consider the TN, LN and GEV EMOS approaches described in Sections \ref{subs:subs3.1.1} -- \ref{subs:subs3.1.3}, respectively, and the raw ensemble and climatological forecasts (observations of the training period are considered as an ensemble) as well. For the sake of brevity here we report only the main results, further details can be found in Appendices \ref{appendix_twcrps} and \ref{appendix_levels}. 

\subsection{Implementation details}
\label{subs:subs4.1}  
In the case studies presented here the estimates of TN and LN EMOS model parameters minimize the mean CRPS of forecast-observation pairs over the training data. Objective functions are optimized using the popular Broyden-Fletcher-Goldfarb-Shanno (BFGS) algorithm \citep[see e.g.][Section 10.9]{press}. However, for the more complex GEV and TGEV models the estimation methods used in the case studies of Sections  \ref{subs:subs4.2} and \ref{subs:subs4.3} differ.
For the short-range forecasts of Section  \ref{subs:subs4.2} we follow the suggestions of \citet{lt13} and calculate the ML estimates of the GEV parameters, whereas for the TGEV model we consider the box constrained version of BFGS \citep[L-BFGS-B;][]{blnz95} and keep the shape parameter \ $\xi$ \ in the interval \ $]-0.278,1/3[$ \ to ensure a finite mean and a positive skewness. Note that the ML estimates of the shape of the GEV model also remain in this interval. For the global ECMWF forecasts of Section  \ref{subs:subs4.2}  both GEV and TGEV parameters are estimated by minimizing the mean CRPS of the training data with the help of a BFGS algorithm, where the constrains on scale and shape parameters are forced using appropriate transformations.
All optimization tasks are performed using the {\tt optim} function of {\tt R} allowing at most 200 iteration steps. In the case of TN and LN models starting parameters of location/mean are computed with a linear regression of the observations on the corresponding forecasts, whereas the starting points for the scale parameters are fixed. In the case of GEV and TGEV models all iterations are started from fixed initial points.

\begin{figure}[t!]
\begin{center}
\includegraphics[width=.9\textwidth]{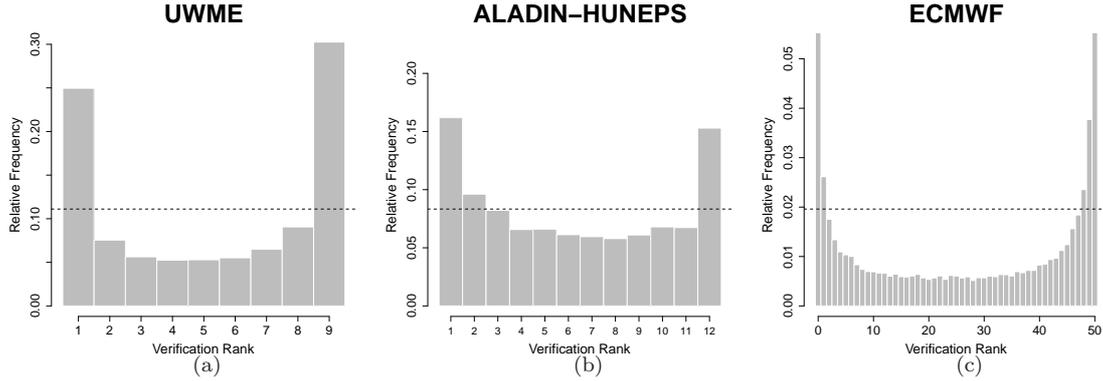}
\vskip -.5 truecm
\centerline{\hbox to 11cm{\scriptsize \qquad (a) \hfill  (b) \hfill (c)}}
\vskip -.3 truecm
\caption{\small Verification rank histograms.  (a) UWME
  for the calendar year 2008; (b) ALADIN-HUNEPS
  ensemble for the period 1 April 2012 -- 31 March 2013; (c) ECMWF ensemble for the period
  1 May 2010 -- 30 April 2011.} 
\label{fig:VRst}
\end{center}
\end{figure}

\subsection{Short-range ensemble forecasts}
  \label{subs:subs4.2}
The case studies of this section are based on those three wind speed data sets that have already been investigated in \citet{bl15,bl16}. We use the same training and verification data for the TGEV modelling (global training with matching training period lengths) as in the earlier works, allowing a direct comparison with the performance of the previously investigated TN, LN and GEV EMOS models.

\subsubsection{EMOS models for the UWME}
\label{subs:subs4.2.1}
As one can observe on Figure \ref{fig:VRst}a, the verification rank histogram of the 8-member UWME wind speed forecasts for calendar year 2008 is highly U-shaped, indicating a strongly underdispersive character. The ensemble range contains the validating observation in only $45.24\%$ of cases, which is far below the nominal coverage of $77.78\%$, calling for some form of calibration.

\begin{table}[t!]
\begin{center}{\small 
\begin{tabular}{lccccc} \hline
Forecast&CRPS&MAE&RMSE&Cover.&Av. w.\\
&$(m/s)$&$(m/s)$&$(m/s)$&$(\%)$&$(m/s)$\\ \hline
TN&1.114 (1.052,1.188)&1.550 (1.466,1.655)&2.048&78.65&4.67 \\
LN&1.114 (1.052,1.188)&1.554 (1.465,1.658)&2.052&77.29&4.69 \\
GEV&1.100 (1.041,1.174)&1.554 (1.463,1.656)&2.047&77.20&4.69 \\
TGEV&1.099 (1.038,1.173)&1.551 (1.464,1.656)&2.046&76.69&4.62 \\\hline
Ensemble&1.353 (1.274,1.460)&1.655 (1.554,1.775)&2.169&45.24&2.53 \\
Climatology&1.412 (1.291,1.539)&1.987 (1.820,2.170)&2.629&81.10&5.90 \\\hline
\end{tabular} 
\caption{\small Mean CRPS and MAE of median forecasts together with $95\,\%$ confidence intervals, RMSE of mean forecasts and coverage and  average width of $77.78\,\%$ central prediction intervals for the UWME. Mean and maximal probability of predicting negative wind speed by the GEV model: $0.05\,\%$ and $4\,\%$.} \label{tab:scoresUWME} 
}
\end{center}
\end{table}

As the 8 members of the UWME are non-exchangeable, for post-processing we make use of TN and LN EMOS models \eqref{eq:tnlocscale} and \eqref{eq:lnlocscale}, respectively, and GEV and TGEV EMOS with parametrization \eqref{eq:gevlocscale}, where \ $K=8$. \ Ensemble forecasts for calendar year 2008 are calibrated using a 30 day training period, which training period length is a result of a detailed preliminary analysis, see \citet{bl15}.

In Table \ref{tab:scoresUWME} a summary of verification scores and coverage and average width of nominal $77.78\,\%$ central prediction intervals are given for the competing EMOS models and the raw and climatological UWME forecasts (27\,481 forecast cases), whereas Table \ref{tab:twcrpssUWME} reports the mean twCRPS values corresponding to various thresholds. Climatological forecasts underperform the raw ensemble in terms of mean CRPS, MAE and RMSE, but have better skill on the tails which is quantified in lower mean twCRPS values. As mentioned before, the underdispersive character of the raw forecasts leads to poor coverage and very sharp central prediction intervals, whereas the climatological prediction intervals are much wider resulting in a far better coverage. EMOS post-processing improves the calibration and forecast skill of the raw ensemble by a wide margin as all EMOS scores but the mean twCRPS corresponding to most extreme wind speeds are much lower than the corresponding scores of raw and climatological forecasts. The advantage in terms of the mean CRPS is significant. The coverage of each calibrated forecast is very close to the nominal value; however, one should also note that these central prediction intervals are less sharp than the intervals calculated from the raw ensemble. From the competing EMOS approaches, the novel TGEV model results in the lowest mean CRPS, RMSE and twCRPS values (which are either identical with or very close to the corresponding GEV EMOS scores), whereas in terms of MAE it is slightly outperformed by the TN EMOS method. Further, the TGEV model leads to the sharpest central prediction intervals, which is naturally connected with a slight decrease in coverage. For a deeper analysis of the tail behaviour of the different EMOS approaches we refer to Figure \ref{fig:twcrpss}a of Appendix \ref{appendix_twcrps} showing the twCRPSS with respect to the TN EMOS as function of the threshold.

\begin{table}[t!]
\begin{center}{\small
\begin{tabular}{lccc} \hline
Forecast&\multicolumn{3}{c}{twCRPS $(m/s)$}\\\cline{2-4}
&$r\!=\!9$&$r\!=\!10.5$&$r\!=\!14$
\\ \hline
TN&0.150 (0.116,0.189)&0.074 (0.054,0.099)&0.010 (0.005,0.016)\\
LN&0.149 (0.115,0.186)&0.073 (0.053,0.098)&0.010 (0.005,0.017)\\
GEV&0.145 (0.112,0.183)&0.072 (0.052,0.095)&0.010 (0.005,0.018)\\
TGEV&0.145 (0.112,0.180)&0.072 (0.052,0.096)&0.010 (0.005,0.017)\\\hline
Ensemble&0.175 (0.134,0.226)&0.085 (0.061,0.115)&0.011 (0.005,0.019)\\
Climatology&0.173 (0.132,0.220)&0.081 (0.058,0.111)&0.010 (0.005,0.017)\\\hline
\end{tabular} 
\caption{\small Mean twCRPS for various thresholds \ $r$ \ together with $95\,\%$ confidence intervals for the
  UWME.} \label{tab:twcrpssUWME} 
}
\end{center}
\end{table}

\begin{figure}[t!]
  \includegraphics[width=\textwidth]{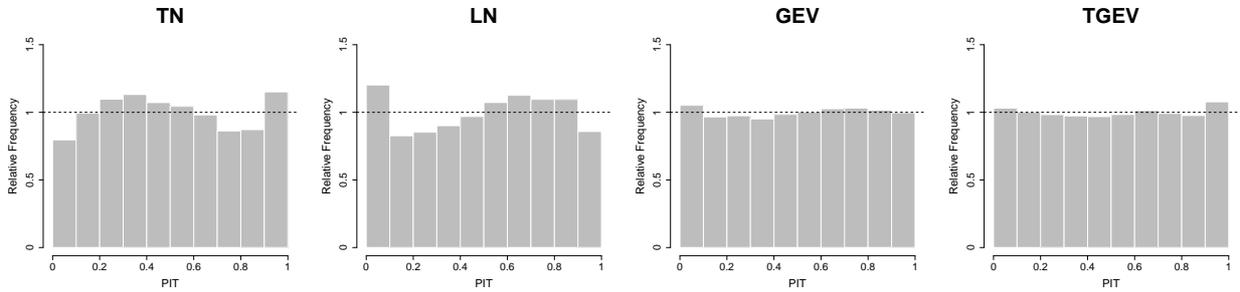}
  \vskip -.5 truecm
   \caption{\small PIT histograms of the EMOS-calibrated UWME forecasts.}
   \label{fig:pitsUWME}
\end{figure}

Finally, compared with the verification rank histogram of the raw UWME forecasts (Figure \ref{fig:VRst}a), the PIT histograms of the different EMOS models displayed in Figure \ref{fig:pitsUWME} are much closer to the desired uniform distribution, indicating an improved calibration. TN and LN EMOS result in slightly biased and hump-shaped  histograms, whereas the histograms of GEV and TGEV approaches are almost perfectly flat. These shapes are nicely in line with the corresponding CRPS values of Table \ref{tab:scoresUWME}.

Based on the above results one can conclude that in the case of the UWME forecasts, from the competing EMOS approaches the novel TGEV model shows the best forecast skill, closely followed by the GEV EMOS. This conclusion is also supported by the results of Appendix \ref{appendix_levels}, where the calibration of the investigated EMOS models at different forecast levels is addressed. However, in connection with the GEV model one should not forget about the positive probability of predicting negative wind speed values. For the UWME forecasts at hand the mean and maximum of these probabilities are $0.05\,\%$ and $4\,\%$, respectively \citep{bl15}.

\subsubsection{EMOS models for the ALADIN--HUNEPS ensemble}
  \label{subs:subs4.2.2}
Compared with the UWME discussed in the previous section, the ALADIN-HUNEPS ensemble is better calibrated.  Although the verification rank histogram given in Figure \ref{fig:VRst}b still shows overconfidence, resulting in large bins at the sides, it is much closer to the uniform distribution than the one in Figure \ref{fig:VRst}a, and the ensemble coverage of $61.21\,\%$ is also closer to the nominal $83.33\,\%$.

\begin{table}[t!]
\begin{center}{\small 
\begin{tabular}{lccccc} \hline
Forecast&CRPS&MAE&RMSE&Cover.&Av.w.\\
&$(m/s)$&$(m/s)$&$(m/s)$&$(\%)$&$(m/s)$\\ \hline
TN&0.738 (0.689,0.793)&1.037 (0.966,1.112)&1.357&83.59&3.53  \\
LN&0.741 (0.690,0.799)&1.038 (0.960,1.125)&1.362&80.44&3.57 \\
GEV&0.737 (0.685,0.793)&1.041 (0.970,1.117)&1.355&81.21&3.54 \\
TGEV&0.736 (0.685,0.793)&1.037 (0.969,1.114)&1.356&82.13&3.53 \\ \hline
Ensemble&0.803 (0.749,0.865)&1.069 (1.001,1.136)&1.373&68.22&2.88 \\
Climatology&1.046 (0.944,1.149)&1.481 (1.333,1.627)&1.922&82.54&4.92 \\\hline 
\end{tabular} 
\caption{\small Mean CRPS and MAE of median forecasts together with $95\,\%$ confidence intervals, RMSE of mean forecasts and coverage and 
  average width of $83.33\,\%$ central prediction intervals for the
  ALADIN-HUNEPS ensemble. Mean and maximal probability of predicting negative wind speed by the GEV model: $0.33\,\%$ and $9.46\,\%$.} \label{tab:scoresALHU} 
}
\end{center}
\end{table}

The structure of the ALADIN-HUNEPS ensemble induces a natural division of the ensemble members into two exchangeable groups: the first contains just the control member, while the second consists of the members obtained from random perturbations of the initial conditions \ ($M=11, \ K=2, \ M_1=1, \ M_2=10$). \ Hence, calibration is performed using EMOS models with distribution locations/means linked to the ensemble members via \eqref{eq:tnlocscaleEx}, \eqref{eq:lnlocscaleEx} and \eqref{eq:gevlocscaleEx}. 

The detailed data analysis of \citet{bhn14} suggests a 43 day training period for EMOS post-processing of ALADIN-HUNEPS ensemble forecasts, leaving 315 calendar days (3\,150 forecast cases) between 15 May 2012 and 31 March 2013 for forecast verification.

\begin{table}[t!]
\begin{center}{\small 
\begin{tabular}{lccc} \hline
Forecast&\multicolumn{3}{c}{twCRPS $(m/s)$}\\\cline{2-4}
&$r\!=\!6$&$r\!=\!7$&$r\!=\!9$\\ \hline
TN&0.102 (0.062,0.147)&0.054 (0.027,0.085)&0.012 (0.003,0.022)\\
LN&0.102 (0.062,0.145)&0.054 (0.028,0.084)&0.011 (0.004,0.022)\\
GEV&0.098 (0.062,0.143)&0.052 (0.026,0.081)&0.011 (0.003,0.021)\\
TGEV&0.099 (0.058,0.145)&0.052 (0.026,0.082)&0.011 (0.003,0.022)\\ \hline
Ensemble&0.112 (0.069,0.163)&0.059 (0.030,0.093)&0.013 (0.004,0.026)\\
Climatology&0.127 (0.076,0.190)&0.064 (0.031,0.102)&0.012 (0.003,0.023)\\\hline 
\end{tabular} 
\caption{\small Mean twCRPS for various thresholds \ $r$ \ together with $95\,\%$ confidence intervals for the ALADIN-HUNEPS ensemble.} \label{tab:twcrpsALHU} 
}
\end{center}
\end{table}

Again, Table \ref{tab:scoresALHU} showing the verification scores of different forecasts and the coverage and average width of nominal $83.33\,\%$ central prediction intervals justifies the use of statistical post-processing. All EMOS models result in reasonably sharp forecasts with coverage values close to the nominal one outperforming both the raw and climatological forecasts in terms of all reported scores. The positive effect of statistical calibration can also be observed on mean twCRPS values provided in Table \ref{tab:twcrpsALHU}; however, one should also be aware of the large uncertainty in the forecasts. Among the different post-processing approaches, the TGEV EMOS yields the lowest mean CRPS and MAE and the sharpest central prediction interval combined with a coverage that is the second closest to the nominal one. However, in terms of twCRPS addressing the predictive performance at high wind speed values, GEV EMOS seems to show better forecast skill. This can also be observed in Figure \ref{fig:twcrpss}b of Appendix \ref{appendix_twcrps}, where the twCRPSS values with respect to the TN EMOS are plotted as function of the threshold. The GEV EMOS clearly outperforms the competitors; however, the situation is nuanced by the fact that in the case of ALADIN-HUNEPS ensemble forecasts the maximal probability of predicting negative wind speed is $9.46\,\%$, and the mean value of these probabilities is also $0.33\,\%$. 

\begin{figure}[t!]
  \includegraphics[width=\textwidth]{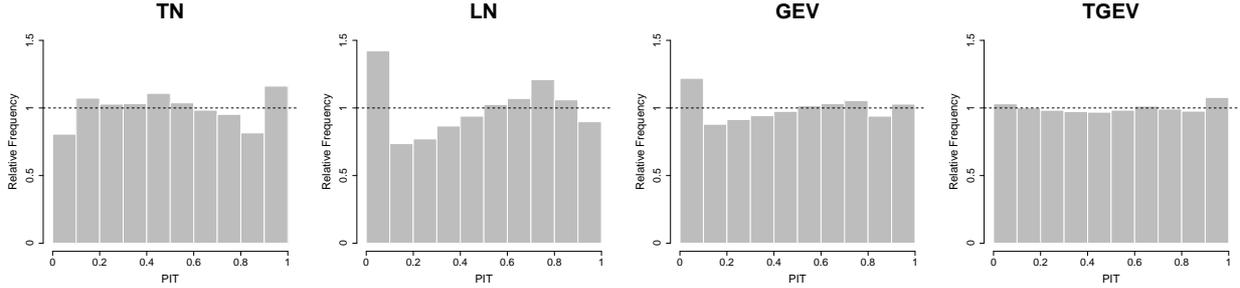}
  \vskip -.5 truecm
    \caption{\small PIT histograms of the EMOS-calibrated ALADIN-HUNEPS ensemble forecasts.}
    \label{fig:pitsALHU}
\end{figure}

The improved calibration of post-processed ALADIN-HUNEPS forecasts can also be observed on PIT histograms of Figure \ref{fig:pitsALHU}, which are much closer to uniformity than the corresponding verification rank histogram, see Figure \ref{fig:VRst}b. Here the TGEV model results in the flattest histogram, whereas the PIT histograms of TN, LN and GEV models are slightly hump-shaped and biased. Hence, keeping in mind also the results of Appendix \ref{appendix_levels}, one can conclude that in the case of the ALADIN-HUNEPS ensemble forecasts, from the presented four EMOS approaches the TGEV has the best overall performance.  

\subsubsection{EMOS models for the ECMWF forecasts for Germany}
  \label{subs:subs4.2.3}
From the three EPSs investigated in Section \ref{subs:subs4.2}, the ECMWF ensemble exhibits the lack of calibration to the highest extent. In most cases the ensemble forecasts either under-, or overestimate the validating observation, resulting in a coverage of $43.40\,\%$, whereas the nominal coverage is $96.08\,\%$. The underdispersive character of the forecasts can also be clearly observed on the corresponding verification rank histogram (see Figure \ref{fig:VRst}c).

The 50 members of operational ECMWF EPS are regarded as exchangeable, so in the link functions \eqref{eq:tnlocscaleEx}, \eqref{eq:lnlocscaleEx} and \eqref{eq:gevlocscaleEx} we have \ $K=1$ \ and \ $\overline{f}_1$ \ equals the ensemble mean. Following the suggestions of \citet{bl15}, the parameters of the EMOS models for calibrating ECMWF ensemble forecast for the period 1 May 2010 -- 30 April 2011 (83\,220 forecast cases) are estimated globally using a rolling training period of length 20 days.

\begin{table}[t!]
\begin{center}{\small
\begin{tabular}{lccccc} \hline
Forecast&CRPS&MAE&RMSE&Cover.&Av.w.\\
&$(m/s)$&
$(m/s)$&$(m/s)$&$(\%)$&$(m/s)$\\ \hline
TN&1.045 (0.974,1.125)&1.388 (1.298,1.488)&2.148&92.19&6.39 \\
LN&1.037 (0.970,1.112)&1.386 (1.298,1.482)&2.138&93.16&6.91 \\
GEV&1.034 (0.960,1.114)&1.388 (1.300,1.488)&2.134&94.84&8.22 \\
TGEV&1.031 (0.962,1.112)&1.385 (1.298,1.480)&2.135&92.89&7.37 \\ \hline
Ensemble&1.263 (1.194,1.345)&1.441 (1.373,1.523)&2.232&45.00&1.80 \\ 
Climatology&1.550 (1.406,1.700)&2.144 (1.948,2.340)&2.986&95.84&11.91 \\ \hline
\end{tabular} 
\caption{\small Mean CRPS and MAE of
  median forecasts together with $95\,\%$ confidence intervals, RMSE of mean forecasts and coverage and average width of $96.08\,\%$ central prediction intervals for the
  ECMWF ensemble forecasts for Germany. Mean and maximal probability of predicting negative wind speed by the GEV model: $0.01\,\%$ and $5\,\%$.} \label{tab:scoresECMWFDE} 
}
\end{center}
\end{table}  
  
\begin{table}[t!]
\begin{center}{\small
\begin{tabular}{lccc} \hline
Forecast&\multicolumn{3}{c}{twCRPS $(m/s)$}\\\cline{2-4}
&$r\!=\!10$&$r\!=\!12$&$r\!=\!15$\\ \hline
TN&0.200 (0.150,0.255)&0.110 (0.075,0.147)&0.042 (0.024,0.062)\\
LN&0.198 (0.146,0.254)&0.109 (0.075,0.149)&0.042 (0.024,0.062)\\
GEV&0.195 (0.145,0.250)&0.106 (0.072,0.145)&0.041 (0.024,0.059)\\
TGEV&0.194 (0.143,0.248)&0.106 (0.072,0.143)&0.041 (0.024,0.060)\\ \hline
Ensemble&0.211 (0.155,0.272)&0.113 (0.077,0.152)&0.043 (0.025,0.061)\\
Climatology&0.251 (0.182,0.326)&0.128 (0.087,0.172)&0.045 (0.026,0.066)\\\hline 
\end{tabular} 
\caption{\small Mean twCRPS for various thresholds \ $r$ \ together with $95\,\%$ confidence intervals for the ECMWF ensemble forecasts for Germany.} \label{tab:twcrpsECMWFDE} 
}
\end{center}
\end{table}
  
Similar to Sections \ref{subs:subs4.2.1} and \ref{subs:subs4.2.2}, in Table \ref{tab:scoresECMWFDE} the mean CRPS, MAE and RMSE of post-processed, raw and climatological forecasts are reported together with the corresponding coverage and average width of $96.08\,\%$ (nominal) central prediction intervals, while Table \ref{tab:twcrpsECMWFDE} provides the mean twCRPS scores for three different thresholds. The picture we get after examining these values is also similar to the previous cases: post-processing results in improved predictive performance and better calibration. The lowest CRPS, MAE and twCRPS values belong to the TGEV EMOS model, which has a fair coverage, but slightly less sharp than the TN and LN EMOS. 

Although the mean twCRPS values and the corresponding $95\,\%$ confidence intervals of GEV and TGEV models given in Table \ref{tab:twcrpsECMWFDE} are almost identical, Figure \ref{fig:twcrpss}c of Appendix \ref{appendix_twcrps} displaying again the twCRPSS with respect to TN EMOS reveals the differences between the tail behaviour of the two methods and indicates the superiority of the novel TGEV EMOS approach. Note also that here the mean and maximal probabilities of predicting negative wind speed by the GEV model are $0.01\,\%$ and $5\,\%$, respectively.
  
\begin{figure}[t!]
  \includegraphics[width=\textwidth]{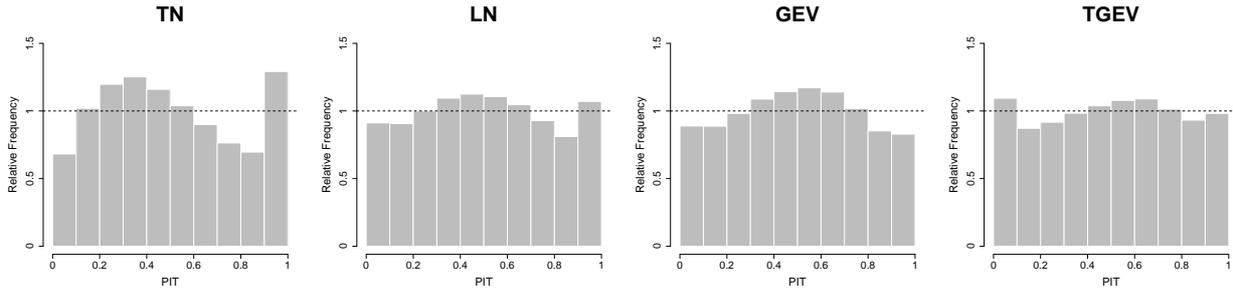}
  \vskip -.5 truecm
    \caption{\small PIT histograms of the EMOS-calibrated ECMWF forecasts for Germany.}
    \label{fig:pitsECMWFDE}
\end{figure}

Finally, the comparison of the PIT histograms of Figure \ref{fig:pitsECMWFDE} with the verification rank histogram of the raw ECMWF ensemble (see Figure \ref{fig:VRst}c) again shows that post-processing substantially improves the calibration of forecasts. However, one should also note that none of the competing EMOS methods results in uniformly distributed PIT values. E.g. the GEV EMOS model is slightly overdispersive having heavy tails, which is fully in line with the wide nominal central prediction intervals (see Table \ref{tab:scoresECMWFDE}), whereas the tails of the TN EMOS model are slightly too light. TGEV and LN EMOS PIT values show the smallest deviation from uniformity, hence, for the studied ECMWF forecasts again the TGEV EMOS model has the best overall performance. Note that this conclusion is rather in line with the corresponding results of Appendix \ref{appendix_levels}.

 \subsection{EMOS models for the global ECMWF ensemble forecasts}
  \label{subs:subs4.3}
  
The case studies of Section \ref{subs:subs4.3} verify the positive effect of EMOS post-processing on calibration of short-term wind speed ensemble forecasts in general, and the superiority of the TGEV EMOS approach as well. However, as argued in the discussion of \citet{frg19}, the longer the lead time, the more training data is needed for post-processing to outperform the raw ensemble, and a similar conclusion can be derived from the results of \citet{bbpbb20}, too. This motivates the case study presented in this section, where calibration of global ECMWF wind speed ensemble forecasts with lead times \ $1,2,\ldots, 15$ \ days covering a very long time period  of almost four and a half years is considered.
  
\begin{figure}[t!]
\centering
\hbox{
\includegraphics[width=\textwidth]{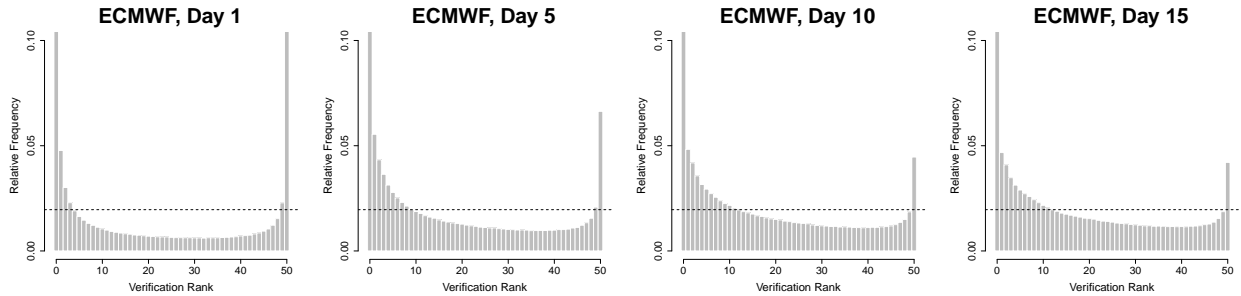}
}
\vskip -.3 truecm
\caption{\small Verification rank histograms of the global ECMWF ensemble forecasts for the period 16 January 2014  -- 25 June 2018.} 
\label{fig:VRHglobal}
\end{figure}  

\begin{figure}[t!]
\begin{center}
\includegraphics[width=.9\textwidth]{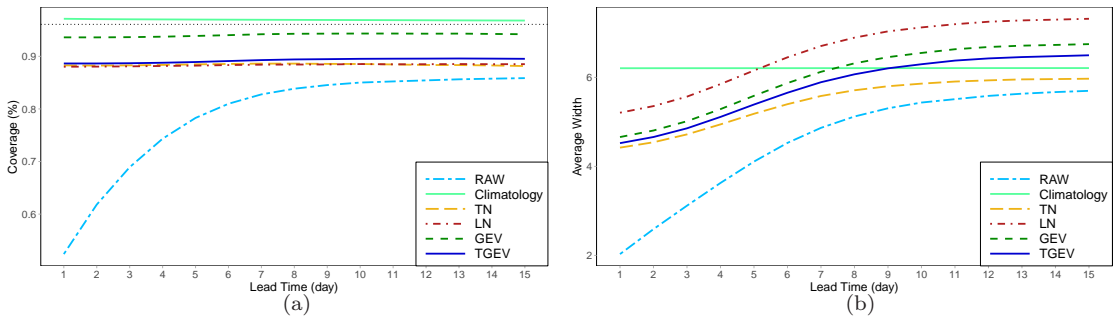}
\vskip -.5 truecm
\centerline{\hbox to 8.5cm{\scriptsize \qquad  (a) \hfill (b) }} 
\vskip -.3 truecm
\caption{\small Coverage (a) and average width (b) of nominal $96.08\,\%$ central prediction intervals. In panel (a) the ideal coverage is indicated by the horizontal dotted line.}
\label{fig:Cov_Avw_ECMWFglobal}
\end{center}
\end{figure}

As one can observe on the verification rank histograms of Figure \ref{fig:VRHglobal}, the global ECMWF forecasts are strongly U-shaped for all lead times; however, the increase of the forecast horizon reduces underdispersion. This might be explained by the increase of forecast uncertainty resulting in wider ensemble range and better coverage, which improves from 52.05\,\% of day 1 to 85.74\,\% of day 15 (see also Figure \ref{fig:Cov_Avw_ECMWFglobal}).

For calibration we use the same EMOS model settings as in Section \ref{subs:subs4.2.3} considering a single group of exchangeable ensemble members; however in this case the large ensemble domain does not allow global modelling. Thus, local estimation with a rolling training period of 100 days is applied, which ensures a reasonably stable parameter estimation for all investigated EMOS approaches and leaves the period 10 May 2014 -- 25 June 2018 (1508 calendar days after excluding the two days with missing data) for validation purposes (1\,596\,972 individual forecast cases for each lead time).

\begin{figure}[t!]
\begin{center}
\includegraphics[width=.9\textwidth]{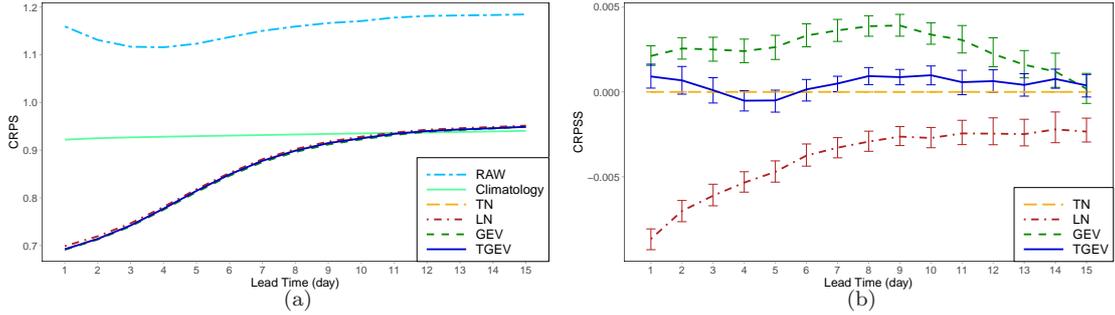}
\vskip -.5 truecm
\centerline{\hbox to 8.5cm{\scriptsize \qquad  (a) \hfill (b) }}
\vskip -.3 truecm
\caption{\small (a) CRPS of the raw, climatological and calibrated ECMWF global forecasts; (b) CRPSS with respect to the TN EMOS model together with 95\,\% confidence intervals.} 
\label{fig:CRPSECMWFglobal}
\end{center}
\end{figure}
  
\begin{table}[b!]
\begin{center}
\resizebox{\textwidth}{!}{
\begin{tabular}{lccccccccccccccc} \hline
Day&1&2&3&4&5&6&7&8&9&10&11&12&13&14&15\\ \hline
Mean &2.48&2.48&2.48&2.49&2.51&2.53&2.54&2.56&2.57&2.59&2.60&2.61&2.63&2.63&2.65\\
Q90&7.36&7.30&7.28&7.32&7.32&7.30&7.30&7.30&7.37&7.42&7.48&7.51&7.58&7.59&7.61\\
Q95&14.20&13.95&13.76&13.59&13.38&13.14&13.02&12.92&13.00&13.02&12.99&13.12&13.22&13.25&13.30\\
Q99&32.95&32.32&31.65&30.82&29.79&29.23&28.53&28.18&27.90&27.83&27.63&27.84&27.84&27.77&27.94 \\ \hline
\end{tabular} 
}
\caption{\small Mean and the 90th, 95th and 99th quantiles of probabilities (in \%) of predicting negative wind
  speed by the GEV model.} 
  \label{tab:GEVProbECMWF} 
\end{center}
\end{table}  

In contrast to the case of ECMWF temperature forecasts investigated in \citet{frg19} or \citet{bbpbb20}, in terms of the mean CRPS all considered EMOS models outperform the raw wind speed ensemble forecasts for all lead times by a wide margin (see Figure \ref{fig:CRPSECMWFglobal}a). Note that the non-monotonic shape of the mean CRPS of the raw ensemble is a result of representativeness error in the verification, which can be partially corrected by adding up observation uncertainty to the ensemble spread \citep{bb20}. For shorter lead times EMOS models are also superior to climatology, but the advantage is decreasing with the lead time and disappears after day 11. To make visible the differences between the various EMOS approaches in terms of the mean CRPS, Figure \ref{fig:CRPSECMWFglobal}b shows the CRPSS values with respect to the TN EMOS model. LN EMOS exhibits the worst forecast skill but the disadvantage decreases with the increase of the forecast horizon. GEV EMOS outperforms its competitors, followed by the TGEV EMOS, which has a significantly positive skill score for almost all lead times. Similar conclusions can be drawn from Figure \ref{fig:regions} of Appendix \ref{appendix_levels} showing the skill scores separately for different forecast levels. However, for this global data set the problem of predicting negative wind speed values by the GEV EMOS approach is far more pronounced than in the case studies of Section \ref{subs:subs4.2}. According to Table \ref{tab:GEVProbECMWF}, the mean of these probabilities is around $2.5\,\%$, whereas the 99th quantiles range from $27.63\,\%$ to $32.95\,\%$, which makes a possible operational use problematic.

\begin{figure}[t!]
\begin{center}
\includegraphics[width=.9\textwidth]{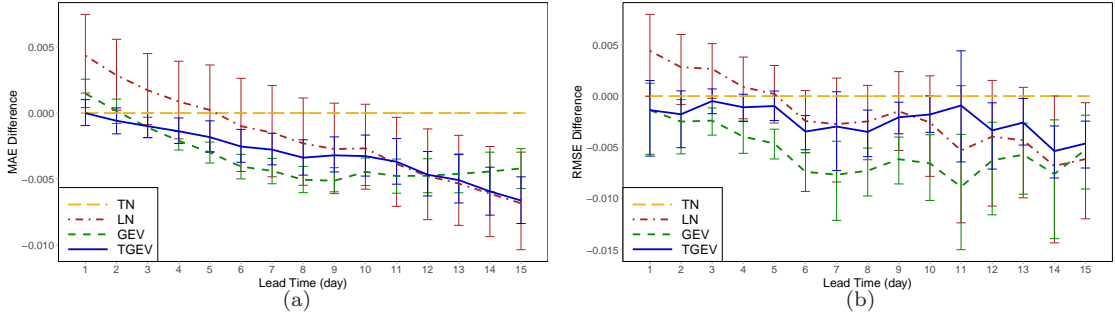}
\vskip -.5 truecm
\centerline{\hbox to 8.5cm{\scriptsize \qquad  (a) \hfill (b) }} 
\vskip -.3 truecm
\caption{\small Difference in MAE (a) and RMSE (b) values from the reference TN EMOS model together with 95\,\% confidence intervals.} 
\label{fig:MAE_RMSE_ECMWFglobal}
\end{center}
\end{figure}
  
In Figures \ref{fig:MAE_RMSE_ECMWFglobal}a,b the differences in MAE and RMSE from the reference TN EMOS model are given (the smaller the better). For short and very long lead times the TGEV EMOS results in the lowest MAE values, whereas between 4 and 10 days the GEV EMOS significantly outperforms its competitors. After day 11 the performance of the LN EMOS is similar to that of the TGEV EMOS; however, the uncertainty of the former is much higher.  A different ranking can be observed in Figure \ref{fig:MAE_RMSE_ECMWFglobal}b, where the GEV EMOS results in the lowest score values, followed by the TGEV EMOS model, which for medium lead times behaves very similarly to the LN EMOS.

As expected, climatological forecasts result in the best coverage (Figure \ref{fig:Cov_Avw_ECMWFglobal}a), closely followed by the GEV EMOS. The coverage values of TGEV, TN and LN EMOS approaches are slightly below $90 \,\%$ for all lead times and the corresponding curves are rather flat and very close to each other. In terms of sharpness, Figure  \ref{fig:Cov_Avw_ECMWFglobal}b shows a clear ranking of the competing post-processing methods for all lead times. TN EMOS results in the narrowest central prediction intervals followed by TGEV, GEV and LN EMOS models.

\begin{figure}[t]
\begin{center}
\hbox{
\includegraphics[width=\textwidth]{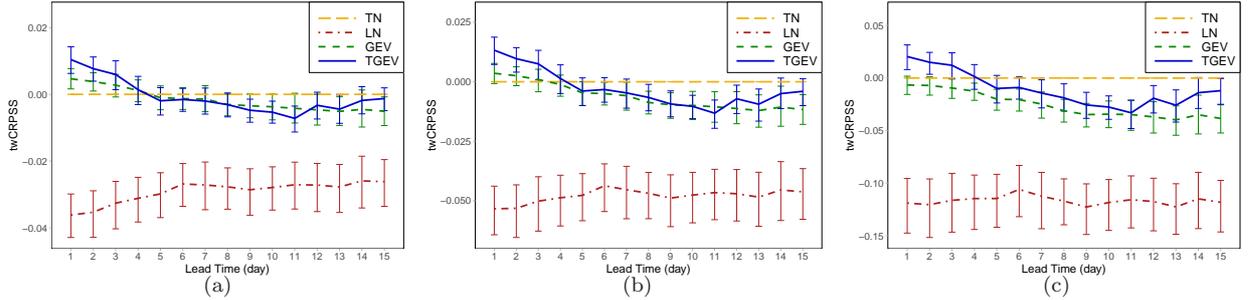}
}
\vskip -.5 truecm
\centerline{\hbox to 12cm{\scriptsize \qquad (a) \hfill  (b) \hfill (c)}}
\vskip -.3 truecm
\caption{\small twCRPSS values with respect to the TN EMOS model for thresholds $6 \ m/s$ (a), $7 \ m/s$ (b) and $9 \ m/s$ (c) together with 95\,\% confidence intervals.} 
\label{fig:twcrpss_ECMWFglobal}
\end{center}
\end{figure}

To compare the tail behaviour of the competing EMOS models we consider the twCRPSS values with respect to the TN EMOS approach for thresholds corresponding again to 90th, 95th and 98th quantiles of the wind speed observations (see Figure \ref{fig:twcrpss_ECMWFglobal}). The ranking of the different EMOS models is consistent for all three investigated thresholds; after day 3 TN EMOS results in the best forecast skill, whereas the LN EMOS approach, similar to Figure \ref{fig:CRPSECMWFglobal}b, is far behind its competitors.

\begin{figure}[t!]
\begin{center}
  \includegraphics[width=\textwidth]{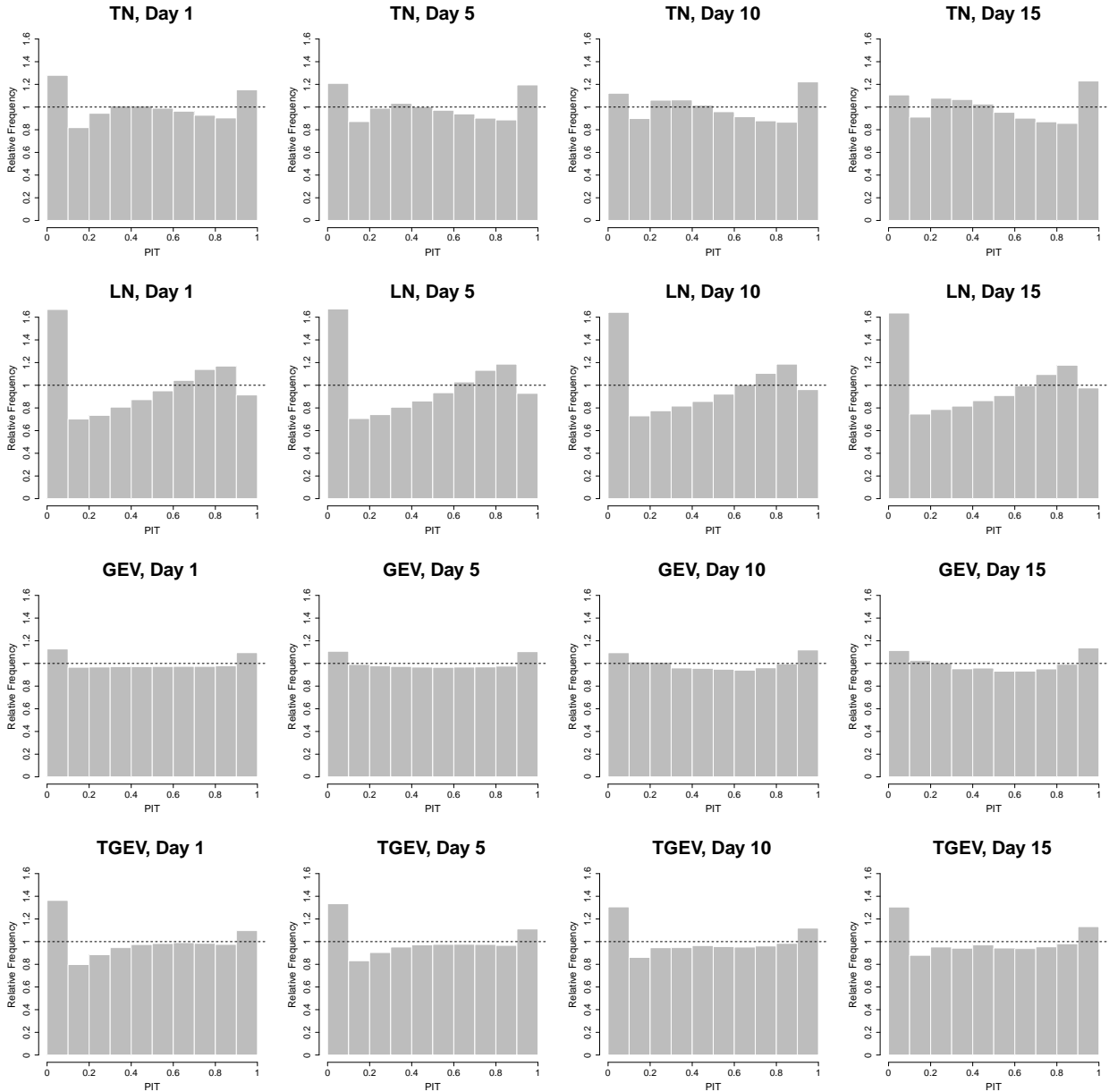}
  \vskip -.3 truecm
\caption{\small PIT histograms of the EMOS post-processed ECMWF global forecasts for days 1, 5, 10 and 15.}
\label{fig:PITglobal}
\end{center}
\end{figure}

Finally, the PIT histograms of EMOS post-processed forecasts for lead times 1, 5, 10 and 15 days plotted in Figure \ref{fig:PITglobal} again show the positive effect of post-processing. They are much closer to uniformity than the verification rank histograms of the raw ECMWF ensemble forecasts of Figure \ref{fig:VRHglobal}; moreover, the shapes of the presented PIT histograms are nicely in line with the corresponding CRPS scores (Figure \ref{fig:CRPSECMWFglobal}) and coverage and average widths of nominal central prediction intervals (Figure \ref{fig:Cov_Avw_ECMWFglobal}). PIT histograms of the LN EMOS approach show the largest deviation from uniformity, whereas the histograms of the GEV model are almost perfectly flat with a slight underdispersion, especially for longer lead times. TGEV EMOS also results in rather flat PIT histograms with slightly light lower tails for all lead times.

For the ECMWF data set at hand the GEV EMOS model shows the best overall predictive performance for all lead times, followed by the TGEV EMOS. However, looking back again to the mean probabilities of predicting negative wind speed by the GEV model given in Table \ref{tab:GEVProbECMWF}, one should prefer the slightly less skillful novel TGEV EMOS approach.

\section{Conclusions}
\label{sec:Conc}
For the purpose of calibrating wind speed ensemble forecasts we propose a novel EMOS approach based on a truncated generalized extreme value distribution. The aim is to correct the deficiency of the efficient GEV EMOS method of \citet{lt13} of occasionally predicting negative wind speed. The TGEV EMOS model is tested both on short-range (24 -- 48 h) wind speed forecasts of three completely different ensemble prediction systems (8-member UWME, 11-member ALADIN--HUNEPS and 50-member ECMWF) covering different and relatively small geographical regions and on a much larger data set of global ECMWF forecasts for four and a half calendar years with lead times from 1 to 15 days. For model verification we use the CRPS of the probabilistic forecasts, the MAE of the median and the RMSE of the mean forecasts, and we also analyze the coverage and the average width of nominal central prediction intervals, which serve as measures of calibration and sharpness, respectively. Further, the predictive performance at high wind speed values is assessed with the help of the twCRPS for thresholds corresponding approximately to the 90th, 95th and 98th percentiles of the observed wind speed. 

The forecast skill of the TGEV EMOS model is compared to that of the TN, LN and GEV EMOS approaches, and the raw and climatological forecasts. According to the results of the presented four case studies, post-processing always improves the calibration of probabilistic and accuracy of point forecasts and all EMOS models outperform both the raw ensemble and climatology. One can also observe that the TGEV EMOS approach has the best overall performance -- regarding the four presented methods -- closely followed by the GEV EMOS model. However, for the latter, at least in the case study of Section \ref{subs:subs4.3}, the mean probability of predicting negative wind speed values is around $2.5\,\%$ for all considered lead times.

In the present study our focus is restricted to univariate forecasts for a single location and lead time. However, most practical applications \citep[e.g. in the context of wind energy forecasting, see][]{pm18} require an accurate modeling of spatial and temporal dependencies. Hence, multivariate extension of the proposed TGEV EMOS model in order to provide spatially and temporally consistent calibrated wind speed forecasts might be an interesting direction of future research. For a detailed overview of the possible approaches see e.g. \citet{lbm20}.

\bigskip
\noindent
{\bf Acknowledgments.} \ S\'andor Baran and Marianna Szab\'o were supported by the ÚNKP-19-3 New National Excellence Program of The Ministry for Innovation and Technology. S\'andor Baran and Patr\'\i cia Szokol were supported by the National Research, Development and Innovation Office under Grant No. NN125679. S\'andor Baran also acknowledges the support of the EFOP-3.6.2-16-2017-00015 project, while Patr\'\i cia Szokol was supported by the EFOP-3.6.1-16-2016-00022 project. Both projects were co-financed by the Hungarian Government and the European Social Fund. The authors are indebted to Zied Ben Bouall\`egue for providing the ECMWF data, the University of Washington MURI group for providing the UWME data and Mih\'aly Sz\H ucs from the HMS for providing the ALADIN-HUNEPS data.

\noindent
\appendix

\section{Mean of a truncated generalized extreme value distribution}
\label{appendix_mean}
To simplify the formulation of the results, similar to the notations of Section \ref{subs:subs3.2}, in what follows we set aside the indication of the parameters of the GEV and TGEV CDFs \ $G$ \ and \ $G_0$ \ defined by \eqref{eq:gevCDF} and \eqref{eq:tgevCDF}, respectively.

The present section is devoted to verification of the formula \eqref{eq:tgevMean} for the TGEV mean in the non-trivial cases when \ $G$ \ and \ $G_0$ \ differ. Let \ $\xi<1$ \ and \ $0<G(0)<1$. \ The PDF \ $g_0(x)$ \ of a \ $\mathcal{TGEV}(\mu,\sigma,\xi)$ \ distribution defined by \eqref{eq:tgevCDF}
equals 
\begin{equation}
    \label{eq:tgev_pdf}
    g_0(x)=\begin{cases}
\frac{\left[1+\xi(\frac{x-\mu}{\sigma})\right]^{-1/\xi-1}\exp\big(-\left[1+\xi(\frac{x-\mu}{\sigma})\right]^{-1/\xi}\big)}{\sigma\left(1-G(0)\right)}, & \quad \text{if \ $\xi\ne 0 $;}\\ 
    \frac{\exp\left(\frac{x-\mu}{\sigma}\right)\exp\left(-\exp\left[-\frac{x-\mu}{\sigma}\right]\right)}{\sigma\left(1-G(0)\right)},& \quad \text{if \ $\xi= 0$,}
    \end{cases}
\end{equation} 
for \ $x\ge 0$ \ and \ $x\xi\ge \mu\xi-\sigma$, \  and \ $g_0(x)=0$ \ otherwise, where 
\begin{equation*}
    G(0)=\begin{cases}
    \exp(-[1-{\xi\mu}/{\sigma}]^{-1/\xi}), & \quad \text{if \ $\xi\ne 0 $,}\\ 
    \exp(-\exp[\mu/\sigma]),& \quad \text{if \ $\xi= 0$.}\\
    \end{cases}
  \end{equation*}

Let \ $X$ \ be a TGEV random variable and assume  \ $\xi\ne 0$ \ and \ $\xi\mu - \sigma \leq 0$. \ If \ $\xi>0$, \ then the support of \ $g_0(x)$ \ is \ $[0,\infty[$, \ so
  \begin{equation}
     \label{eq:integral}
    {\mathsf E}X=\frac{1}{\sigma(1-G(0))}\int_0^{\infty} x \left[1+\xi\left(\frac{x-\mu}{\sigma}\right)\right]^{-1/\xi-1}\exp\left(-\left[1+\xi\left(\frac{x-\mu}{\sigma}\right)\right]^{-1/\xi}\right)\mathrm{d} x.
\end{equation}
 For \ $\xi<0$ \ the support of \ $g_0(x)$ \ changes to \ $[0,\mu-\sigma/\xi]$, \  so the integral in \eqref{eq:integral} should be taken over this particular interval. However, in both cases the change of variables leads to
\begin{align*}
     {\mathsf E}X&=\frac{1}{1\!-\!G(0)}\int_{0}^{\left(1-\frac{\xi\mu}{\sigma}\right)^{-1/\xi}} \left[\frac{(t^{-{\xi}}-1)\sigma}{\xi}+\mu\right]\exp(-t)\mathrm{d} t \\
     &=\mu - \frac{\sigma}{\xi}+ \frac{\sigma(\Gamma_{\ell}(1-\xi,[1-{\xi\mu}/{\sigma}]^{-1/\xi}))/\xi }{1-\exp (-[1-{\xi\mu}/{\sigma}]^{-1/\xi})}.
\end{align*}

Finally, let \ $\xi=0$. \ In this case
\begin{equation*}
   {\mathsf E}X= \frac{1}{\sigma(1-G(0))}\int_{0}^{\infty} x \exp\left(\frac{x-\mu}{\sigma}\right)\exp\left(-\exp\left[-\frac{x-\mu}{\sigma}\right]\right) \mathrm{d} x,
\end{equation*}
where the change of variables with respect to  \ $t=\exp\left(-\frac{x-\mu}{\sigma}\right)$ \ results in
\begin{equation*}
    {\mathsf E}X=\frac{1}{\sigma(1-G(0))}\int\limits_{0}^{\exp(\mu/\sigma)}  (\mu-\sigma\ln t) \exp(-t)\mathrm{d} t=
    \frac{\mu + \sigma(C-\ei(-\exp[{\mu}/{\sigma}]))}{1-\exp(-\exp[\mu/\sigma])}.
\end{equation*}
\proofend

\section{CRPS of a truncated generalized extreme value distribution}
\label{appendix_crps}
Following the ideas of \citet{ft12}, the CRPS of a TGEV distribution is derived using representation
\begin{equation}\label{eq:CRPSquant}
\crps(G_0,x)=x\big(2G_0(x)-1\big)-2\int_0^1 t G_0^{-1}(t){\mathrm d} t+ 2\int_{G_0(x)}^1 G_0^{-1}(t){\mathrm d} t,
\end{equation}
where \ $G_0^{-1}$ \ denotes the quantile function corresponding to \ $G_0$. \  Short calculation shows that for \ $0<y<1$ \ 
\begin{equation*}
    G_0^{-1}(y)=\begin{cases}
    \mu+\frac{\sigma}{\xi}\Big(-1+\big[-\ln \tau(y)\big]^{-\xi}\Big), &  \text{if \ $\xi\ne 0$,}\\
    \mu-\sigma\Big(\ln\big[-\ln \tau(y)\big]\Big),&   \text{if \ $\xi=0$,}
  \end{cases}    \quad \text{where} \quad \tau(y):=\big(1-G(0)\big)y+G(0).
\end{equation*}
Assume first \ $\xi\ne 0$. \ Then the first integral of \eqref{eq:CRPSquant} equals
\begin{align*}
   2\int_0^1 t G_0^{-1}(t){\mathrm d} t&=\mu-\frac{\sigma}{\xi}+\frac{2\sigma}{\xi}\!\int_0^1\! t\big[\!-\ln \tau(t)\big]^{-\xi}{\mathrm d} t=\mu-\frac{\sigma}{\xi}+\frac{2\sigma}{\xi}\!\int_{G(0)}^1\! \frac{\tau\!-\!G(0)}{(1\!-\!G(0))^2}[-\ln \tau ]^{-\xi}{\mathrm d} \tau \\
   &=\mu-\frac{\sigma}{\xi}+\frac{2\sigma}{\xi} \frac 1{(1-G(0))^2}\Bigg[\int_{G(0)}^1 \tau [-\ln \tau ]^{-\xi}{\mathrm d} \tau    - G(0)\int_{G(0)}^1 [-\ln \tau ]^{-\xi}{\mathrm d} \tau\Bigg].
\end{align*}
Now, let \ $\Gamma_u$ \ denote the upper incomplete gamma functions, defined as
\begin{equation*}
 \Gamma_u(a,x)=\int_x^{\infty} t^{a-1}{\mathrm e}^{-t} {\mathrm d} t. \qquad 
\end{equation*}
Using  \ $\Gamma(a)=\Gamma_{\ell}(a,x)+\Gamma_u(a,x)$, \ short calculations involving appropriate changes of variables show
\begin{align*}
   \int_{G(0)}^1 \tau [-\ln \tau ]^{-\xi}{\mathrm d} \tau &=
    2^{\xi-1}\Big[\Gamma (1-\xi)-\Gamma_u\big(1-\xi,-2\ln G(0)\big)\Big]=2^{\xi-1}\Gamma_{\ell}\big(1-\xi,-2\ln G(0)\big),\\ 
   \int_{G(0)}^1 [-\ln \tau ]^{-\xi}{\mathrm d} \tau &=
    \Gamma (1-\xi)-\Gamma_u\big(1-\xi,-\ln G(0)\big)=\Gamma_{\ell}\big(1-\xi,-\ln G(0)\big).
\end{align*}
Hence,
\begin{align}
   \label{eq:CRPSint1}
   2\int_0^1 t G_0^{-1}(t){\mathrm d} t&=\mu-\frac{\sigma}{\xi}+\frac{\sigma} {\xi(1-G(0))^2}\Big[2^{\xi}\Gamma_{\ell}\big(1-\xi,-2\ln G(0)\big)   - G(0)\Gamma_{\ell}\big(1-\xi,-\ln G(0)\big)
   \Big].
\end{align}
The second integral of \eqref{eq:CRPSquant} can be evaluated in a similar way, resulting in
\begin{align}
   \label{eq:CRPSint2}
  \int_{G_0(x)}^1 G_0^{-1}(t){\mathrm d} t =\big(1-G_0(x)\big)\Big(\mu-\frac{\sigma}{\xi}\Big)+\frac{\sigma}{\xi(1-G(0))}\Gamma_{\ell}\big(1-\xi,-\ln G(x))\big).
 \end{align}
Finally, the combination of equations \eqref{eq:CRPSquant}, \eqref{eq:CRPSint1} and \eqref{eq:CRPSint2} gives
\begin{align*}
    \crps(G_0,x)=\big(2G_0(x)&\,-1\big)\Big(x-\mu+\frac{\sigma}{\xi}\Big)+
    \frac{\sigma}{\xi(1-G(0))^2}\Big[-2^{\xi}\Gamma_{\ell}\big(1-\xi,-2\ln G(0)\big) \\
    & +2G(0)\Gamma_{\ell}\big(1-\xi,-\ln G(0)\big)+2\big(1-G(0)\big)\Gamma_{\ell}\big(1-\xi,-\ln G(x)\big) \Big].
\end{align*}

Now, let \ $\xi=0$. \ In this case for the integrals in \eqref{eq:CRPSquant} we have
\begin{align*}
   2\int_0^1 t G_0^{-1}(t){\mathrm d} t&=\mu-2\sigma \int_0^1 t \ln\big[-\ln \tau(t)\big]\mathrm{d}t=\mu-2\sigma \int_{G(0)}^1 \frac{\tau-G(0)}{(1-G(0))^2}\ln [-\ln \tau ]{\mathrm d} \tau \\
   &=\mu-\frac{2\sigma}{(1-G(0))^2}\Bigg[\int_{G(0)}^1 \tau \ln [-\ln \tau ]{\mathrm d} \tau    - G(0)\int_{G(0)}^1 \ln [-\ln \tau ]{\mathrm d} \tau\Bigg], \\
     \int_{G_0(x)}^1 G_0^{-1}(t){\mathrm d} t &=\mu\big(1-G_0(x)\big)-\sigma \int_{G_0(x)}^1 \ln\big[-\ln \tau(t)\big]\mathrm{d}t \\ 
     &=\mu\big(1-G_0(x)\big)-\frac{\sigma}{1-G(0)} \int_{G(x)}^1 \ln\big[-\ln \tau\big]\mathrm{d}\tau.
\end{align*}
Hence, keeping in mind that
$$\int\!\tau \ln\!\big[\!-\ln \tau\big]\mathrm{d}\tau \!=\!\frac{\tau^2}2\ln\!\big[\!-\ln \tau\big]\!-\!\frac 12\ei(2\ln \tau)\big] \ \ \text{and} \ \ \int\!\ln\!\big[\!-\ln \tau\big]\mathrm{d}\tau \!=\!\tau\ln\!\big[\!-\ln \tau\big]\!-\!\ei(\ln \tau)\big],$$
we obtain
\begin{align*}
    \crps(G_0,&\,x)\!=\!x(2G_0(x)\!-\!1)\!+\!\mu\!-\!2\mu G_0(x)\!+\!\frac{2\sigma}{(1\!-\!G(0))^2}\Bigg\{\bigg[\frac{s^2}{2}\ln\!\big[\!-\ln s\big]\!-\!\frac 12\ei(2\ln s)\bigg]_{s=G(0)}^{s=1}\\
    &-G(0)\Big[(s\ln\!\big[\!-\ln s\big]\!-\!\ei(\ln s ) \Big]_{s=G(0)}^{s=1} \!-\!
    \big(1\!-\!G(0)\big)\Big[s\ln\!\big[\!-\ln s\big]\!-\!\ei(\ln s)\Big]_{s=G(x)}^{s=1}\Bigg\}.
\end{align*}
Finally, since 
\begin{align*}
    s^2\ln &\big[\!-\ln s\big]-\!\ei(2\ln s )\!-\!2G(0)\big(s\ln\!\big[\!-\ln s\big]\!-\!\ei(\ln s)\big)\!-\!2\big(1\!-\!G(0)\big)\big(s\ln\!\big[\!-\ln s\big] \!-\!\ei(\ln s)\big) \\
    &=s^2\ln\big[\!-\ln s\big]-2s\ln\big[\!-\ln s\big]-\ei(2\ln s )+2\ei(\ln s )\\
    &=C-\ln 2 +(s-1)^2\ln\big[\!-\ln s\big]+\sum\limits_{k=1}^{\infty}\frac{-(2\ln s )^k+2(\ln s )^k}{k!k} \to C-\ln 2 \quad \text{as} \quad s \uparrow 1,
\end{align*}
the CRPS of a TGEV distribution with \ $\xi=0$ \ equals
\begin{align*}
    \crps(G_0,x)=&\,(x-\mu)\big(2G_0(x)-1\big) +\frac{\sigma}{(1-G(0))^2} \\
    &\times\Big(C-\ln 2+
    \ei\big(2\ln G(0)\big)+(G(0))^2\ln\big[-\ln G(0)\big]-
    2G(0)\ei\big(\ln G(0)\big)\Big)\\
    &+\frac{2\sigma}{1-G(0)}\Big[G(x)\ln\big[-\ln G(x)\big]-\ei\big(\ln G(x)\big) \Big].
\end{align*}
\proofend

\section{Dependence of twCRPS of short-range forecasts on the threshold}
\label{appendix_twcrps}

\begin{figure}[t!]
\begin{center}
  \hbox{\includegraphics[width=0.48\textwidth]{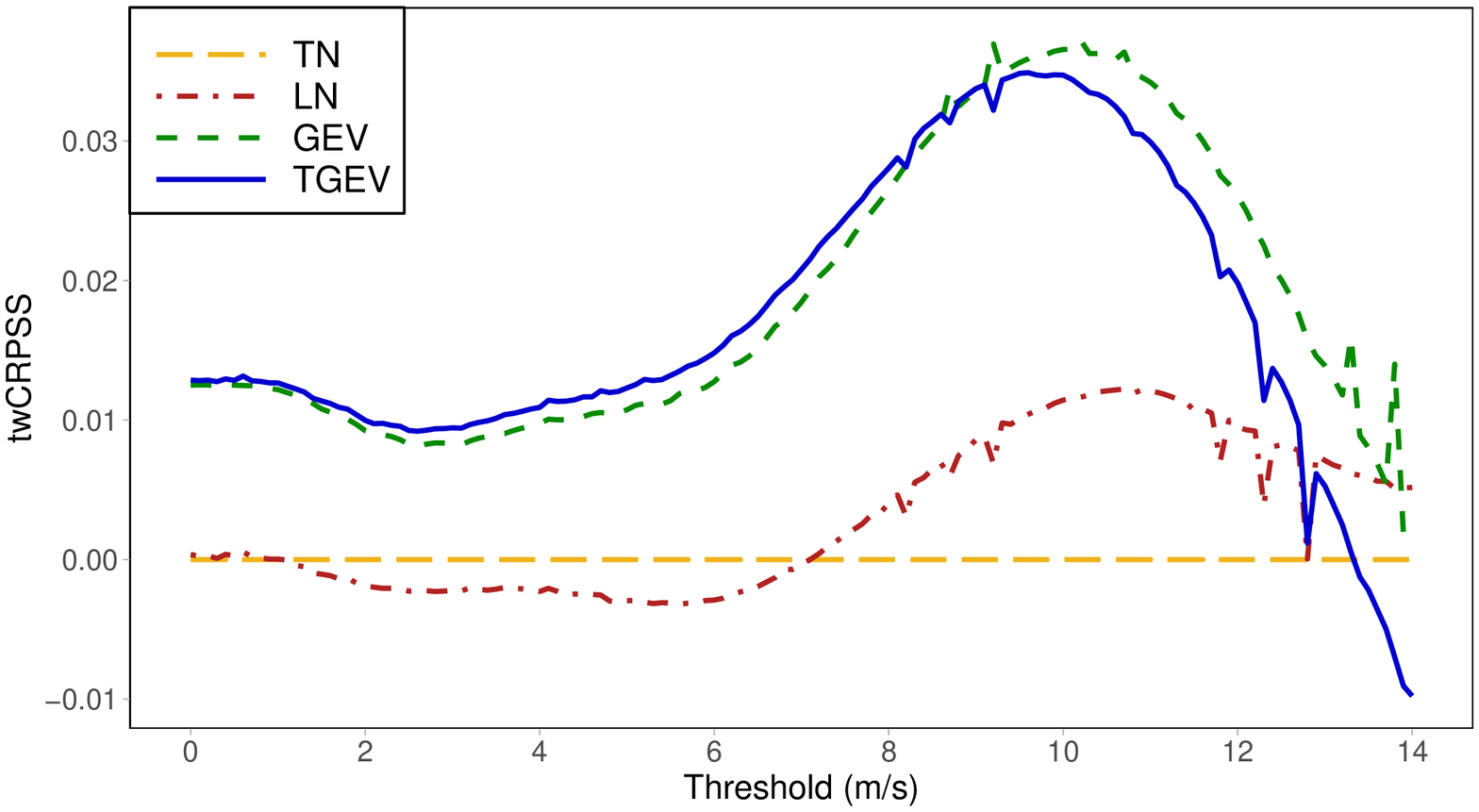} \
   \includegraphics[width=0.48\textwidth]{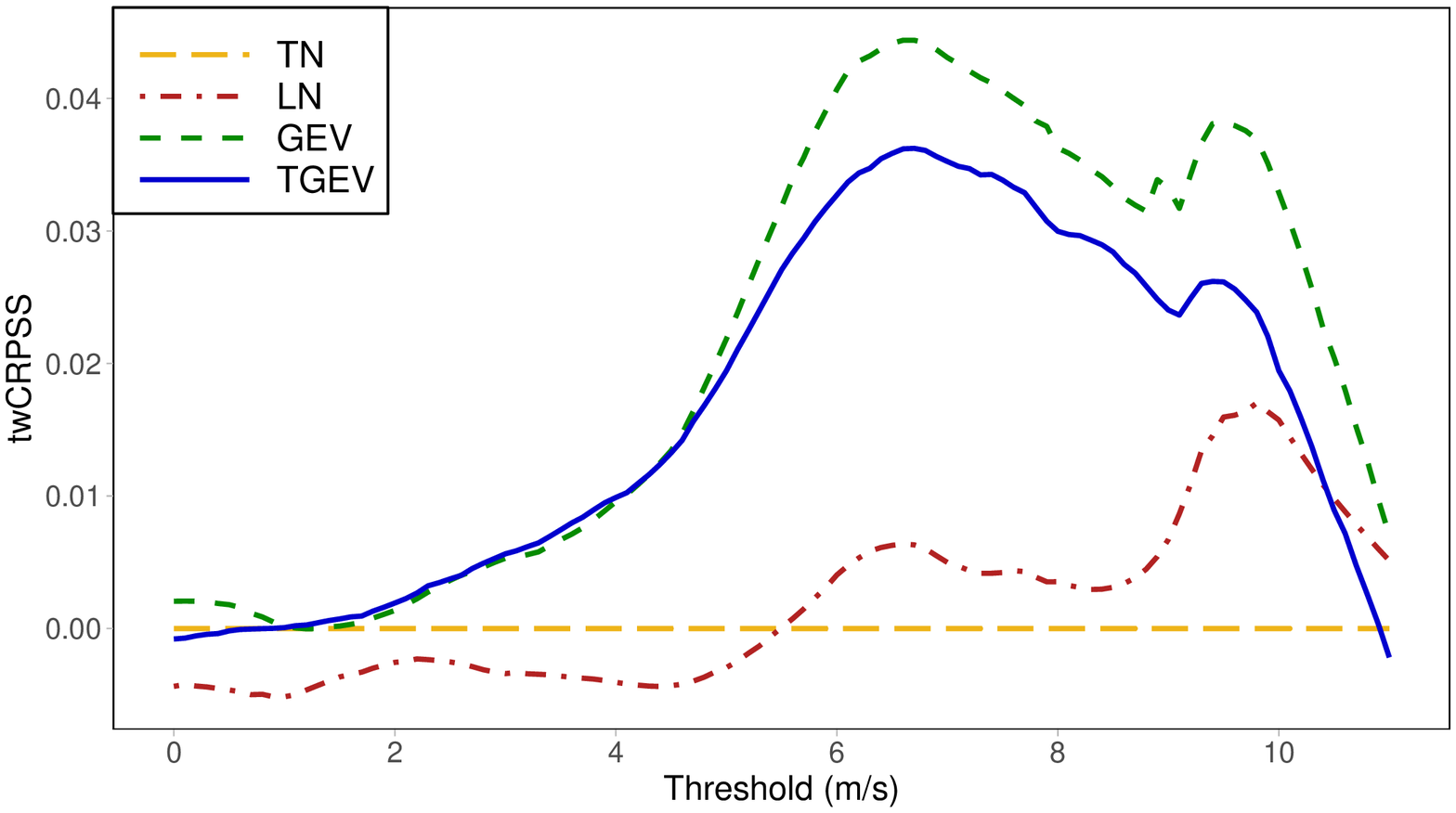}}
  \vskip -.5 truecm
\centerline{\hbox to 8.5cm{\scriptsize \qquad  (a) \hfill (b) }} 
 \vskip .2 truecm
\includegraphics[width=0.5\textwidth]{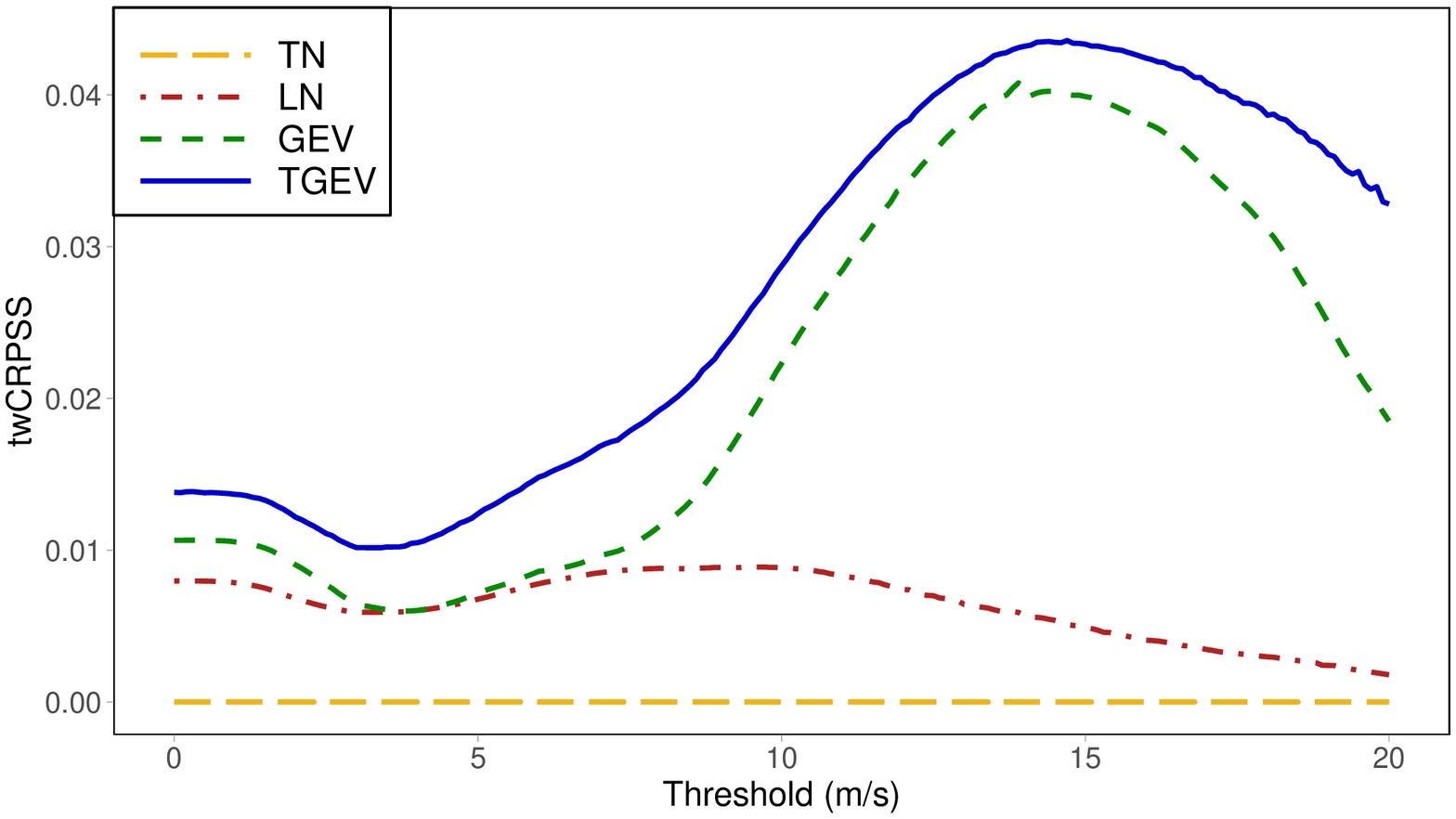}
 \vskip -.5 truecm
\centerline{\scriptsize \quad (c)} 
\vskip -.3 truecm
     \caption{\small twCRPSS values with respect to the TN EMOS model. (a) UWME; (b) ALADIN-HUNEPS ensemble; (c) ECMWF ensemble.}
   \label{fig:twcrpss}
\end{center}
\end{figure}

Beyond comparing the twCRPS values reported in Tables \ref{tab:twcrpssUWME}, \ref{tab:twcrpsALHU} and \ref{tab:twcrpsECMWFDE}, one can get a deeper insight into the tail behaviour of the different EMOS approaches by examining Figure \ref{fig:twcrpss} showing the twCRPSS with respect to the TN EMOS as function of the threshold. For the UWME forecasts (Figure \ref{fig:twcrpss}a), GEV and TGEV models show very similar behaviour and up to 13 m/s both approaches outperform the TN and LN EMOS methods. For lower threshold values TGEV EMOS results in the highest skill score, but after 8 m/s GEV shows the best predictive performance. A similar ranking of the methods can be observed in Figure \ref{fig:twcrpss}b; however, here the interval where the GEV and TGEV methods perform almost identically is much shorter. Finally, in the case of the ECMWF ensemble the TGEV EMOS results in the highest skill score for all thresholds, see Figure \ref{fig:twcrpss}c.

The difference between the first two cases and the third one in terms of the GEV and TGEV EMOS models might be explained with the difference in the support of these distributions. In the case of UWME and ALADIN-HUNEPS forecasts, the shape parameter $\xi$ is negative for all forecast cases. Hence, the supports of GEV and TGEV predictive distributions are \ $]-\infty, \mu - \sigma /\xi]$ \ and \ $[0,\mu - \sigma /\xi]$, \ respectively, moreover, the upper bounds of the GEV are in general higher than those of the TGEV. In this way GEV can capture higher wind speeds, which results in better forecast skill in the upper tail.  However, for the ECMWF ensemble the shape parameter of GEV and TGEV distributions is positive in $99.18\,\%$ and $92.88\,\%$ of all forecast cases, respectively,  meaning that in these cases the supports of the predictive distributions are not bounded from above.

\section{Calibration at different forecast levels}
\label{appendix_levels}

\begin{table}[t!]
\begin{center}{\small 
    \begin{tabular}{lccccccccc} \hline
Forecast&\multicolumn{3}{c}{UWME}&\multicolumn{3}{c}{ALADIN-HUNEPS} &\multicolumn{3}{c}{ECMWF} \\ \cline{2-10}
&Low&Medium&High&Low&Medium&High&Low&Medium&High\\ \hline
LN&0.000&0.000&-0.000&-0.031&-0.004&0.003&0.013&0.008&0.006\\
GEV&0.035&0.008&0.025&0.005&-0.001&0.018&0.027&0.008&0.016\\
TGEV&0.034&0.008&0.028&0.008&-0.001&0.020&0.019&0.010&0.028\\\hline
\end{tabular} 
\caption{\small CRPSS values with respect to the TN EMOS model for forecasts with low (less than the 10th percentile), medium (between 10-90 percentiles) or high (greater than the 90th percentile) ensemble mean.} \label{tab:regions} 
}
\end{center}
\end{table}

\begin{figure}[t!]
\begin{center}
  \includegraphics[width=\textwidth]{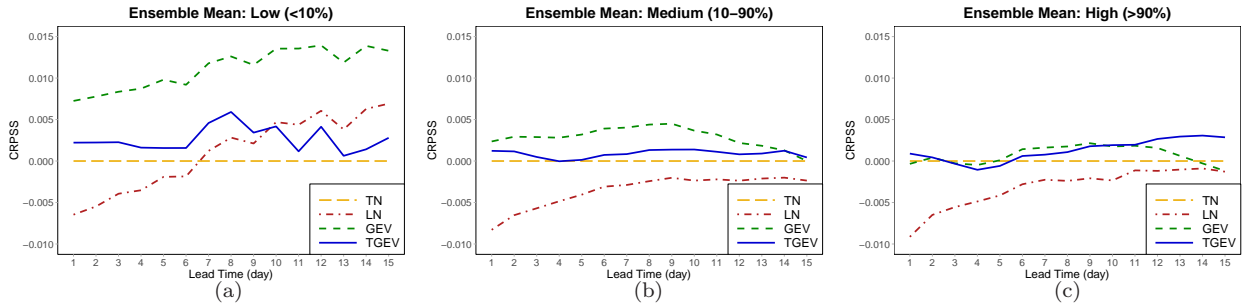}
  \vskip -.5 truecm
\centerline{\hbox to 12cm{\scriptsize \qquad (a) \hfill  (b) \hfill (c)}}
\vskip -.3 truecm
     \caption{\small CRPSS values with respect to the TN EMOS model for forecasts with low (a), medium (b) or high (c) ensemble mean.}
   \label{fig:regions}
\end{center}
\end{figure}

To investigate calibration at different forecast levels, using the idea of \citet{brem19}, we group the forecast cases of verification data according to whether the ensemble mean is low (less than the 10th percentile of the means for the given lead time), medium (between 10-90 percentiles) or high (greater than the 90th percentile). As in terms of the mean CRPS all calibrated forecasts outperform the raw ensemble by a wide margin in all of our case studies, here we focus on the comparison of the competing EMOS approaches. Table \ref{tab:regions} contains the CRPSS values with respect to the reference TN EMOS model at different levels for the short-range forecasts of Section \ref{subs:subs4.2}. Note that for low and high forecasts the GEV and TGEV EMOS approaches outperform both the TN and LN EMOS models and the TGEV EMOS shows the best overall performance. 

A different behaviour can be observed in Figure \ref{fig:regions} showing the same skill score as function of the lead time for the global ECMWF forecasts investigated in Section \ref{subs:subs4.3}. Here the GEV EMOS is the overall winner; however, for high wind speed forecasts the differences between the various EMOS approaches are rather small, especially for long forecast horizons. Note also that for the medium and high groups the ranking of the models is consistent with Figure \ref{fig:CRPSECMWFglobal}b.

\begin{thebibliography}{99}
{\small 
\bibitem[Baran {\em et al.\/}, 2020]{bbpbb20} Baran, S., Baran, \'A.,
  Pappenberger, F. \& Ben Bouall\`egue, Z. (2020). Statistical
  post-processing of heat index ensemble forecasts: is there a royal road?
  {\em Quarterly Journal of the Royal Meteorological Society,\/} 146,
  3416--3434.
  
\bibitem[Baran {\em et al.\/}, 2014]{bhn14} Baran, S., Hor\'anyi, A. \&
  Nemoda, D. (2014). Comparison of the BMA and EMOS statistical methods in
  calibrating temperature and wind speed forecast ensembles.
  {\em Id\H oj\'ar\'as,\/} 118, 217--241.

\bibitem[Baran and Lerch, 2015]{bl15} Baran, S. \&  Lerch, S. (2015).
  Log-normal distribution based Ensemble Model Output Statistics models for
  probabilistic wind speed forecasting. {\em Quarterly Journal of
    the Royal Meteorological Society,\/} 141, 2289--2299.
    
\bibitem[Baran and Lerch, 2016]{bl16} Baran, S. \&  Lerch, S. (2016). 
  Mixture EMOS model for calibrating ensemble forecasts of wind speed. 
  {\em Environmetrics,\/} 27, 116--130.

\bibitem[Baran and Lerch, 2018]{bl18} Baran, S. \& Lerch, S. (2018). Combining
  predictive distributions for the statistical post-processing of ensemble
  forecasts. {\em International Journal of Forecasting,\/} 34, 477--496.
  
\bibitem[Bassetti {\em et al.\/}, 2018]{bcr18} Bassetti, F., Casarin, R. \&
  Ravazzolo, F. (2018). Bayesian nonparametric calibration and combination of
  predictive distributions. {\em Journal of the American Statistical
    Association,\/} 113, 675--685.
  
\bibitem[Ben Bouall\`egue {\em et al.\/}, 2013]{btg13} Ben Bouall\`egue, Z.,
  Theis, S. E. \& Gebhardt, C. (2013). Enhancing COSMO-DE ensemble forecasts by
  inexpensive techniques.  {\em Meteorologische Zeitschrift,\/} 22, 49--59.  
  
\bibitem[Ben Bouall\`egue, 2020]{bb20} Ben Bouall\`egue, Z. (2020). Accounting
  for representativeness in the verification of ensemble forecasts. {\em ECMWF
    Technical Memorandum,\/} No. 865, doi:10.21957/5z6esc7wr.  

\bibitem[Bremnes, 2019]{brem19} Bremnes, J. B. (2019). Constrained quantile
  regression splines for ensemble postprocessing. {\em Monthly Weather
    Review,\/} 147, 1769--1780.

\bibitem[Buizza, 2018]{buizza18} Buizza, R. (2018). Ensemble forecasting and 
	the need for calibration. In Vannitsem, S., Wilks, D. S., Messner, J. W.
     (Eds.), {\em Statistical Postprocessing of Ensemble Forecasts\/},
     Elsevier, Amsterdam, pp. 15--48.
     
\bibitem[Buizza  {\em et al.\/}, 2005]{bhtp05} Buizza, R., Houtekamer, P. L.,
  Toth, Z., Pellerin, G., Wei, M. \& Zhu, Y. (2005). A comparison of the ECMWF,
  MSC, and NCEP global ensemble prediction systems.  {\em Monthly Weather
    Review,\/} 133, 1076--1097.    
  
\bibitem[Byrd {\em et al.\/}, 1995]{blnz95} Byrd, R. H., Lu, P., Nocedal, J.
  \& Zhu, C. (1995). A limited memory algorithm for bound constrained
  optimization. {\em SIAM Journal on Scientific Computing,\/} 16, 1190--1208.   

\bibitem[Dawid, 1984]{dawid84} Dawid, A. P. (1984). Present position and
  potential developments: Some personal views: Statistical theory: The
  prequential approach. {\em Journal of the Royal Statistical Society. Series
    A (Statistics in Society),\/} 147, 278--292.

\bibitem[Descamps {\em et al.\/}, 2015]{dljbac}  Descamps, L., Labadie, C.,
  Joly, A., Bazile, E., Arbogast, P. \& C\'ebron, P. (2015). PEARP, the
  M\'et\'eo-France short-range ensemble prediction system. {\em Quarterly
    Journal of the Royal Meteorological Society,\/} 141, 1671--1685.
 
\bibitem[Feldmann {\em et al.\/}, 2019]{frg19} Feldmann, K., Richardson,
  D. S. \& Gneiting, T. (2019). Grid‐ versus station-based postprocessing of
  ensemble temperature forecasts. {\em Geophysical Research Letters,\/} 46,
  7744--7751.
  
\bibitem[Friederichs and Hense, 2007]{fh07} Friederichs, P. \& Hense, A.
  (2007). Statistical downscaling of extreme precipitation events using
  censored quantile regression. {\em Monthly Weather Review,\/} 135,
  2365--2378.  
  
\bibitem[Friederichs and Thorarinsdottir, 2012]{ft12} Friederichs, P. \&
 Thorarinsdottir, T. L. (2012). Forecast verification for extreme value
 distributions with an application to probabilistic peak wind prediction.
 {\em Environmetrics,\/} 23, 579--594.
  
\bibitem[Garcia {\em et al.\/}, 1998]{gtpd98} Garcia, A., Torres, J. L.,
  Prieto, E. \& De Francisco, A. (1998). Fitting wind speed distributions: A
  case study. {\em Solar Energy,\/} 62, 139--144.
  
\bibitem[Gneiting, 2011]{gneiting11} Gneiting, T. (2011). Making and
  evaluating point forecasts. {\em Journal of the American Statistical Association,\/} 106, 746--762.
    
\bibitem[Gneiting, 2014]{gneiting14} Gneiting, T. (2014). Calibration of
  medium-range weather forecasts. {\em ECMWF Technical Memorandum,\/} No. 719, 
  doi:10.21957/8xna7glta.

\bibitem[Gneiting and Raftery, 2005]{gr05} Gneiting, T. \& Raftery, A. E.
  (2005). Weather forecasting with ensemble methods. {\em Science,\/} 310,
  248--249.

\bibitem[Gneiting and Raftery, 2007]{gr07} Gneiting, T. \& Raftery,
  A. E. (2007). Strictly proper scoring rules, prediction and
  estimation. {\em Journal of the American Statistical Association,\/} 102,
  359--378.  

\bibitem[Gneiting {\em et al.\/}, 2005]{grwg05} Gneiting, T., Raftery, A. E.,
  Westveld, A. H. \& Goldman, T. (2005). Calibrated probabilistic forecasting
  using ensemble model output statistics and minimum CRPS estimation. {\em
    Monthly Weather Review,\/} 133, 1098--1118.

\bibitem[Gneiting and Ranjan, 2011]{gr11} Gneiting, T. \& Ranjan, R. (2011).
  Comparing density forecasts using threshold- and  quantile-weighted scoring
  rules. {\em Journal of Business \& Economic Statistics,\/} 29, 411--422.     
    
\bibitem[Gneiting and Ranjan, 2013]{gr13} Gneiting, T. \& Ranjan, R. (2013).
  Combining predictive distributions. {\em Electronic Journal of Statistics,\/}
  7, 1747--1782.    
    
\bibitem[Good, 1952]{good52} Good, I. J. (1952). Rational decisions.
  {\em Journal of the Royal Statistical Society. Series B (Statistical
    Methodology),\/} 14, 107--114.  
    
\bibitem[Grell {\em et al.\/}, 1995]{grell95} Grell, G. A., Dudhia, J. \&
  Stauffer, D. R. (1995) A description of the fifth-generation Penn state/NCAR
  mesoscale model (MM5), NCAR Tech. Note, NCAR/TN-398+STR, Boulder,
  122.

\bibitem[Hor\'anyi {\em et al.\/}, 2006]{hkkr06} Hor\'anyi, A., Kert\'esz, S.,
  Kullmann, L. \& Radn\'oti, G. (2006). The ARPEGE/ALADIN mesoscale numerical
  modeling system and its application at the Hungarian Meteorological Service.
  {\em Id\H oj\'ar\'as,\/} 110, 203--227.

\bibitem[Iversen {\em et al.\/}, 2011]{iversen11} Iversen, T., Deckmin, A.,
  Santos, C., Sattler, K., Bremnes, J. B., Feddersen, H. \&  Frogner, I.-L.
  (2011). Evaluation of 'GLAMEPS' -- a proposed multimodel 
  EPS for short range forecasting. {\em Tellus A,\/} 63, 513--530.

\bibitem[Justus {\em et al.\/}, 1978]{jhmg78} Justus, C. G., Hargraves, W. R.,
  Mikhail, A. \& Graber D. (1978). Methods for estimating wind speed frequency
  distributions. {\em Journal of Applied Meteorology,\/} 17, 350--353.

\bibitem[Leith, 1974]{lei74} Leith, C. E. (1974). Theoretical skill of
  Monte-Carlo forecasts.  {\em Monthly Weather Review,\/} 102, 409--418.

\bibitem[Lerch and Baran, 2017]{lb17} Lerch, S. \& Baran, S. (2017). 
  Similarity-based semi-local estimation of EMOS models. {\em Journal of the
    Royal Statistical Society. Series C (Applied Statistics),\/} 66, 29--51.

\bibitem[Lerch {\em et al.\/}, 2020]{lbm20} Lerch, S., Baran, S., M\"oller, A.,
  Gro\ss, J., Schefzik, R., Hemri, S. \&  Graeter, M. (2020). Simulation-based
  comparison of multivariate ensemble post-processing methods.
  {\em Nonlinear Processes in Geophysics,\/}  27, 349--371.
  
\bibitem[Lerch and Thorarinsdottir, 2013]{lt13} Lerch, S. \&
  Thorarinsdottir, T. L. (2013). Comparison of non-homogeneous
  regression models for probabilistic wind speed forecasting. {\em
    Tellus A,\/} 65, 21206.

\bibitem[Leutbecher and Palmer, 2008]{lp08} Leutbecher, M. \& Palmer,
  T. N. (2008). Ensemble forecasting. {\em Journal of Computational Physics,\/}
  227, 3515--3539.   

\bibitem[Molteni {\em et al.\/}, 1996]{mbp96} Molteni, F., Buizza, R. \&
  Palmer, T. N. (1996). The ECMWF ensemble prediction system: methodology and
  validation. {\em Quarterly Journal of the Royal Meteorological Society,\/}
  122, 73--119.

\bibitem[Murphy, 1973]{murphy73} Murphy, A. H. (1973). Hedging and skill
  scores for probability forecasts. {\em Journal of Applied Meteorology,\/} 12,
  215--223.

\bibitem[National Weather Service, 1998]{nws98} National Weather Service
  (1998). Automated Surface Observing System (ASOS) users guide. National
  Weather Service: Silver Spring, MD.
  https://www.weather.gov/media/asos/aum-toc.pdf. Accessed 5 September
  2020.

\bibitem[Pinson and Messner, 2018]{pm18} Pinson, P. \& Messner, J. W. (2018).
  Application of postprocessing for renewable energy. In Vannitsem, S., Wilks,
  D. S., Messner, J. W. (Eds.), {\em Statistical Postprocessing of Ensemble
    Forecasts\/},  Elsevier, Amsterdam, pp. 241--266.

\bibitem[Politis and Romano, 1994]{pr94} Politis, D. N. \& Romano, J. P. (1994).
  The stationary bootstrap. {\em Journal of the American Statistical
    Association,\/} 89, 1303--1313.  
  
\bibitem[Press {\em et al.\/}, 2007]{press} Press, W. H., Teukolsky, S. A.,
  Vetterling, W. T. \& Flannery, B. T. (2007). {\em Numerical
    Recipes 3rd Edition: The Art of Scientific Computing.\/} Cambridge
  University Press, Cambridge.

\bibitem[Raftery {\em et al.\/}, 2005]{rgbp05} Raftery, A. E., Gneiting, T.,
  Balabdaoui, F. \& Polakowski, M. (2005). Using Bayesian model averaging to
  calibrate forecast ensembles. {\em Monthly Weather Review,\/} 133, 1155--1174.

\bibitem[Rasp and Lerch, 2018]{rl18} Rasp, S. \& Lerch, S. (2018). Neural
  networks for postprocessing ensemble weather forecasts. {\em Monthly Weather
    Review,\/} 146, 3885--3900.
    
\bibitem[Scheuerer, 2014]{sch14} Scheuerer, M. (2014). Probabilistic
  quantitative precipitation forecasting using ensemble model output
  statistics. {\em Quarterly Journal of the Royal Meteorological Society,\/}
  140, 1086--1096.    
  
\bibitem[Sloughter {\em et al.\/}, 2010]{sgr10} Sloughter,
  J. M., Gneiting, T. \& Raftery, A. E. (2010).  Probabilistic wind
  speed forecasting using ensembles and Bayesian model averaging. {\em
    Journal of the American Statistical Association,\/} 105, 25--37.   
    
\bibitem[Taillardat and Mestre, 2020]{tm20} Taillardat, M. \& Mestre, O.
  (2020). From research to applications -- examples of operational ensemble
  post-processing in France using machine learning. {\em Nonlinear Processes
    in Geophysics,\/} 27, 329--347.
    
\bibitem[Thorarinsdottir and Gneiting, 2010]{tg10} Thorarinsdottir, T. L. \&
  Gneiting, T. (2010). Probabilistic forecasts of wind speed: Ensemble model
  output statistics by using heteroscedastic censored regression. {\em Journal
    of the Royal Statistical Society. Series A (Statistics in Society),\/} 173,
  371--388.

\bibitem[Wilks, 2011]{w11} Wilks, D. S. (2011). {\em Statistical Methods in the
    Atmospheric Sciences. 3rd ed.\/} Elsevier, Amsterdam.

\bibitem[Wilks, 2018]{wilks18} Wilks, D. S. (2018). Univariate ensemble
  forecasting. In Vannitsem, S., Wilks, D. S., Messner, J. W. (Eds.), {\em
    Statistical Postprocessing of Ensemble Forecasts\/}, Elsevier, pp. 49--89.  

\bibitem[Yuen {\em et al.\/}, 2018]{emos} Yuen, R. A., Baran, S., Fraley,
  C., Gneiting, T., Lerch, S., Scheuerer, M. \& Thorarinsdottir, T. L. (2018)
  {\em R package ensembleMOS, Version 0.8.2: Ensemble Model Output
    Statistics.\/} https://cran.r-project.org/package=ensembleMOS. Access\-ed 5
  September 2020.
}
\end{thebibliography}
\end{document}